\documentclass[fleqn,usenatbib]{mnras}

\usepackage{newtxtext,newtxmath}

\usepackage[T1]{fontenc}
\usepackage[nolist]{acronym}

\newcommand\stu{\texttt{ST-U}}
\newcommand\multinest{\texttt{MultiNest}}

\begin{acronym}
\acro{1D}{one dimensional}
\acro{AMP}{accreting millisecond pulsar}
\acro{ART-XC}[\textit{ART-XC}]{\textit{Astronomical Roentgen Telescope X-ray Concentrator}}
\acro{cosi}[$\cos i$]{cosine of the observer inclination}
\acro{CDF}{cumulative distribution function}
\acro{CI}{credible interval}
\acro{d}[$D$]{distance}
\acro{GW}{gravitational waves}
\acro{GPU}{graphics processing unit}
\acro{EoS}{equation of state}
\acro{eXTP}[\textit{eXTP}]{\textit{enhanced X-ray Timing and Polarimetry}}
\acro{IXPE}[\textit{IXPE}]{\textit{Imaging X-ray Polarimetry Explorer}}
\acro{XMM-Newton}[\textit{XMM-Newton}]{\textit{X-ray Multi-Mirror Newton}}
\acro{i}[$i$]{observer inclination}
\acro{J1444}{SRGA J144459.2$-$604207}
\acro{kch}{kilo core-hour}
\acro{KDE}{Kernel Density Estimation}
\acro{LMXB}{low-mass X-ray binary}
\acrodefplural{LMXB}{low-mass X-ray binaries}
\acro{m}[$M$]{mass}
\acro{MAP}{maximum a posteriori}
\acro{MP}{millisecond pulsar}
\acro{nh}[$N_{\rm H}$]{neutral hydrogen column density}
\acro{NICER}[NICER]{Neutron star Interior Composition Explorer}
\acro{NS}{neutron star}
\acro{obsid}{Observation ID}
\acro{phi}[$\phi$]{phase}
\acro{prior}{prior probability distribution}
\acro{posterior}{posterior probability distribution}
\acro{PPM}{pulse-profile modelling}
\acro{PRE}{photospheric radius expansion}
\acro{r}[$R_{\rm eq}$]{equatorial radius}
\acro{rin}[$R_{\rm in}$]{inner disc radius}
\acro{RMP}{rotation powered millisecond pulsar}
\acro{SRG}[\textit{SRG}]{Spectrum-Roentgen-Gamma}
\acro{STU}[\texttt{ST-U}]{Single Temperature - Unshared}
\acro{tau}[$\tau$]{optical depth}
\acro{tbb}[$T_{\rm seed}$]{seed photon temperature}
\acro{TBO}{thermonuclear burst oscillation}
\acro{te}[$T_{\rm e}$]{electron temperature}
\acro{theta}[$\theta$]{co-latitude}
\acro{tc}[$\theta_{\rm c}$]{the constant parameters}
\acro{tin}[$T_{\rm in}$]{inner disc temperature}
\acro{X-PSI}[\texttt{X-PSI}]{X-ray Pulse Simulation and Inference}
\acro{zeta}[$\zeta$]{angular radius}
\end{acronym}

\DeclareRobustCommand{\VAN}[3]{#2}
\let\VANthebibliography\thebibliography
\def\thebibliography{\DeclareRobustCommand{\VAN}[3]{##3}\VANthebibliography}

\usepackage{graphicx}	
\usepackage{amsmath}	
\usepackage{orcidlink}

\title[Pulse profile modelling with disc occultation]{Pulse profile modelling of accreting millisecond pulsars with disc occultation and its impact on parameter inference}

\author[Y.-H. Mao et al.]{
Ying-Han Mao\,\orcidlink{0000-0001-8356-2233},$^{1,2}$\thanks{maoyinghan03@gmail.com} 
Bas Dorsman\,\orcidlink{0000-0002-9407-0733},$^{3}$
Anna L. Watts\,\orcidlink{0000-0002-1009-2354},$^{3,4}$
Tuomo Salmi\,\orcidlink{0000-0001-6356-125X},$^{5}$
Juri Poutanen\,\orcidlink{0000-0002-0983-0049},$^{6}$ 
\newauthor
and Xiang-Dong Li\,\orcidlink{0000-0002-0584-8145}$^{1,2}$ \thanks{lixd@nju.edu.cn} 
\\
$^{1}$ School of Astronomy and Space Science, Nanjing University, Nanjing 210023, P. R. China \\
$^{2}$ Key Laboratory of Modern Astronomy and Astrophysics (Nanjing University), Ministry of Education, Nanjing 210023, P. R. China \\
$^{3}$ Anton Pannekoek Institute for Astronomy, University of Amsterdam, Science Park 904, 1098XH Amsterdam, The Netherlands \\
$^{4}$ Gravitation and Astroparticle Physics Amsterdam (GRAPPA), University of Amsterdam, 1098XH Amsterdam, The Netherlands \\
$^{5}$ Department of Physics, P.O. Box 64, FI-00014 University of Helsinki, Finland \\
$^{6}$ Department of Physics and Astronomy, FI-20014 University of Turku, Finland}

\date{Accepted XXX. Received YYY; in original form ZZZ}

\pubyear{\the\year{}}

\begin{document}
\label{firstpage}
\pagerange{\pageref{firstpage}--\pageref{lastpage}}
\maketitle

\begin{abstract}
Pulse profile modelling is a relativistic ray-tracing technique used to infer neutron star mass, radius, and surface hotspot properties from X-ray pulsations. Pulse profile modelling has been widely applied to rotation-powered millisecond pulsars, where the local environment is relatively empty. Application to accreting millisecond pulsars is complicated by the geometry of the local accretion flow, including disc occultation of surface emission. In this work, we extend an established pulse profile modelling code, X-PSI, to incorporate accretion disc occultation in accreting millisecond pulsar pulse profile modelling. We quantify how disc occultation depends on system geometry and evaluate its impact on parameter inference. We find that disc occultation is primarily governed by the viewing inclination and can significantly reshape pulse profiles at moderate to high inclinations. Using synthetic Neutron Star Interior Composition Explorer datasets, we investigate parameter recovery for two representative hotspot configurations. For hotspots close to the rotational poles, statistically acceptable fits can yield posteriors that deviate noticeably from the true parameters. In contrast, in a case with hotspots located closer to the equator we find more reliable parameter recovery. We further find that neglecting disc occultation can introduce spurious posterior modes with comparable statistical support, potentially affecting the interpretation of inferred neutron star parameters, suggesting that this effect should be included in accreting millisecond pulsar pulse profile modelling.
\end{abstract}

\begin{keywords}
accretion, accretion discs -- equation of state -- stars: neutron -- X-rays: binaries
\end{keywords}



\section{Introduction}

Pulse profile modelling (PPM) is a relativistic ray-tracing technique used to model pulsed emission from X-ray emitting hot spots on the surfaces of neutron stars  \citep[NS; see e.g.][]{Pechenick83,Miller1998,Poutanen2003,Morsink2007,Bogdanov19b,Bogdanov21}.  It yields information on the NS mass and radius, which depend on the dense matter equation of state, a topic of intense investigation \citep{chatziioannou25}. It also provides insight into the properties of the X-ray emitting regions.  For rotation-powered millisecond pulsars, sources for which PPM has been enabled by data from  the Neutron Star Interior Composition Explorer  \citep[NICER;][]{Gendreau16}, this has delivered mass-radius measurements at the $\sim$ 10\% level and maps of the heated magnetic polar caps \citep[see e.g][]{Salmi24,Dittmann24,Choudhury24,Mauviard25, Kini26}.

PPM can also be applied to accreting NSs, in particular accreting millisecond pulsars \citep[AMPs;][]{Poutanen2003,Salmi18}.  For these sources the magnetic field of the star is strong enough to truncate the inner edge of the accretion disc some distance from the stellar surface \citep{Patruno21,DiSalvo22}.  Surface hot spots are formed as material from the inner edge of the accretion disc is channeled onto the magnetic poles \citep[e.g.][]{Romanova2004}.  AMPs are interesting targets for PPM because they exhibit multiple independent observable phenomena that depend on mass and radius \citep[such as the properties of thermonuclear bursts and burst oscillations, see, for example,][]{Nattila2017,Galloway2024,Kini24}, allowing us to address modelling systematics. In addition, radiation scattered in the accretion column becomes polarized \citep{Viironen04}; this provides additional information on the system geometry, helping to break degeneracies \citep{Poutanen20}.  Polarized emission from an AMP has now been observed for the first time \citep{Papitto2025}, using data from the \textit{Imaging X-ray Polarimetry Explorer} \citep[\textit{IXPE};][]{IXPE2022}. 

As part of this effort, the X-ray Pulse Simulation and Inference (\acs{X-PSI}) package \citep{Riley23}, which has been used extensively in PPM studies of rotation-powered millisecond pulsars, has been adapted to support PPM for AMPs. This has involved the implementation of an appropriate atmosphere model \citep{Bobrikova2023} and the capability to simulate polarized radiation \citep{Salmi21,Salmi25}. Parameter recovery tests using a number of synthetic data sets have been carried out \citep{Dorsman2025,Salmi25}. \citet{Dorsman26} used the \acs{X-PSI} pipeline for a PPM study of the AMP SAX J1808.4$-$3658 (hereafter J1808), using NICER data from its 2019 and 2022 outbursts.

One additional new aspect that needs to be considered when modelling AMPs is the role of the accretion disc. The disc has several important effects, including emitting X-rays, blocking X-rays from the NS surface, reflecting and reprocessing incident radiation. The first of these was studied in \citet{Dorsman2025,Dorsman26} who considered a geometrically thin accretion disc emitting a multi-temperature blackbody spectrum and also producing a pulse-phase-independent background component that could be marginalised over in the inference.  In this paper, we explore the second role of the disc, occulting the NS surface emission.

In the standard framework of \citet{Shakura1973}, the accretion disc is geometrically thin and optically thick over a broad range of radii. Under these conditions, the emergent radiation can be approximated as a (modified) blackbody with a radially varying effective temperature. This leads to the well-known multicolour disc description \citep{Mitsuda1984,Makishima1986}, which has been widely used also in spectral modelling of AMPs \citep[e.g.][and references therein]{Illiano2023,Li2023,Malacaria2025}. This assumption is also adopted in \citet{Dorsman2025}. In such a configuration, photons emitted from the NS surface along the trajectory intersecting the disc are absorbed rather than transmitted, leading to geometric obscuration of the surface emission.

Previous work has incorporated accretion disc occultation into PPM and directly compared theoretical predictions with observations. Models that include the blocking of one emitting spot (in a configuration where there are two antipodal spots) by the accretion disc have been used to explain several observed properties of J1808 \citep{Poutanen2009, Ibragimov2009,Kajava2011}. The same disc occultation framework was later applied by \citet{Molkov2024}, who showed that the pulse profiles of SRGA J144459.2$-$604207 (hereafter J1444) can be well reproduced when disc occultation is taken into account. More recently, \citet{Dorsman2026} employed this framework in PPM of NICER and IXPE observations of J1444, using the disc occultation model developed in this work. From a broader perspective, disc obscuration of X-ray emission has also been invoked in non-AMP systems such as Her X-1 \citep{Petterson1991,Scott2000}. These studies demonstrate that disc occultation can significantly affect observed pulse profiles and therefore should be treated in PPM.

The inclusion of disc occultation is a physically self-consistent component of PPM in AMPs. Accreted material from the companion star carries angular momentum and is expected to form an accretion disc around the NS. In accreting states, the inner disc radius $R_{\rm in}$ is expected to lie within the corotation radius 
\begin{equation}
\label{eq:rco}
R_{\rm co} = 31~{\rm km}~(M/1.4{\rm M}_\odot)^{1/3}(f/400~{\rm Hz})^{-2/3} ,
\end{equation}
where $M$ is the NS mass and $f$ is the pulsar rotational frequency.
Systems with $R_{\rm in} > R_{\rm co}$ enter the propeller regime which suppresses accretion \citep{Illarionov1975,Rappaport2004, Romanova2005, Papitto2015}.  
For typical rotation rates of AMPs, $R_{\rm co}\simeq$20--40~km and thus is comparable to the NS radius. At such radii, the inner accretion disc can geometrically obscure hotspots located on the southern hemisphere of the NS. 

Motivated by these considerations, we extend \acs{X-PSI} to include disc occultation and evaluate its impact on pulse profiles and parameter inference, enabling the incorporation of more comprehensive physical processes into full Bayesian inference pipelines and facilitating their application to real observational data. We test the model using synthetic data and quantify the impact of disc occultation on both the pulse profiles and the recovery of source parameters, in order to assess whether its inclusion has a measurable effect. This paper is organized as follows. Section~\ref{sec:modelling} introduces PPM for AMPs and outlines the physical description for including disc obscuration. Section~\ref{sec: Effect of disc occultation across parameter space} quantifies the strength of disc occultation across the relevant parameter space. Section~\ref{sec:Parameter inference with disc occultation} presents a parameter inference analysis to evaluate the impact of disc occultation on recovered model parameters. Section~\ref{sec:discussion} discusses the implications of these results, and Section~\ref{sec:conclusion} summarises the main conclusions of this work.

\section{Modelling}\label{sec:modelling}

In this work, we use \acs{X-PSI}, the open-source package for modelling NS X-ray pulse profiles and inferring model parameters. We extend this framework by incorporating accretion disc occultation into the PPM. In the following, we briefly summarise the model adopted for AMPs and describe the physical structure of the disc occultation model.

\subsection{PPM for AMPs}
In PPM of AMPs, we consider two components: the NS surface and the accretion disc. 
To construct the energy-dependent beaming function of the surface radiation, we adopt the accretion shock model of \citet{Bobrikova2023} who approximated it as a homogeneous plane-parallel slab of hot electrons Comptonizing blackbody radiation from the bottom. 
This model is characterized by three parameters: the seed photon blackbody temperature $T_{\rm bb}$, the temperature $T_{\rm e}$ of the electron gas and the Thomson optical depth $\tau$ across the slab, which controls the probability of scattering and is a dominant factor determining the angular distribution of the escaping radiation. 
Using this surface emission model, we compute the propagation of photons in the oblate Schwarzschild approximation, accounting for relativistic light bending around a rapidly rotating NS \citep{Miller1998,Poutanen2003,Morsink2007,AlGendy2014}.
The observer inclination $i$ is defined as the angle between the spin axis and the line of sight of the observer, which is restricted to be smaller than $90\degr$. For each emitting surface element, we determine whether it is visible to the observer \citep[see e.g.][for details]{Bogdanov19b}. The observed pulse profile is then obtained by integrating the contributions from all visible photon trajectories as a function of rotational phase.

We adopt the \stu{} hotspot geometry, consisting of two circular, single-temperature hotspots with unshared parameters. The two hotspots are distinguished by an ordering in colatitude: the hotspot with the smaller colatitude is defined as the primary hotspot, with parameters denoted by the subscript `p', while the other is referred to as the secondary hotspot, with parameters denoted by the subscript `s'. Each hotspot is characterized by three geometric parameters: the colatitude $\theta_{\rm p}$ ($\theta_{\rm s}$), the angular radius $\zeta_{\rm p}$ ($\zeta_{\rm s}$), and the rotational phase $\phi_{\rm p}$ ($\phi_{\rm s}$).

For the second component, the accretion disc, we consider two effects: emission and occultation. The intrinsic disc emission is described by the \texttt{diskbb} model \citep{Mitsuda1984,Makishima1986}, which computes the total disc contribution by integrating the emission from individual rings from the inner to the outer disc radius. The free parameters of this model are the inner disc radius $R_{\rm in}$ and the temperature at the inner radius $T_{\rm in}$. Further details of the disc emission and the modelling of AMPs described above can be found in \citet{Dorsman2025}.

All emission components are subject to absorption by the interstellar medium during their propagation to the observer. The attenuation of the X-ray flux is described as $F_{\rm obs}(E) = F_{\rm int}(E)\ {\rm e}^{-\sigma(E) N_{\rm H}}$, 
where $F_{\rm int}(E)$ and $F_{\rm obs}(E)$ are the intrinsic and observed fluxes at photon energy $E$, and $\sigma(E)$ is the energy-dependent photoelectric cross section. 
This attenuation is implemented using the \texttt{tbnew} model \citep{Wilms2000}, with $N_{\rm H}$ as a free parameter.

\subsection{Criterion for disc occultation}
\begin{figure}
	\includegraphics[width=\columnwidth]{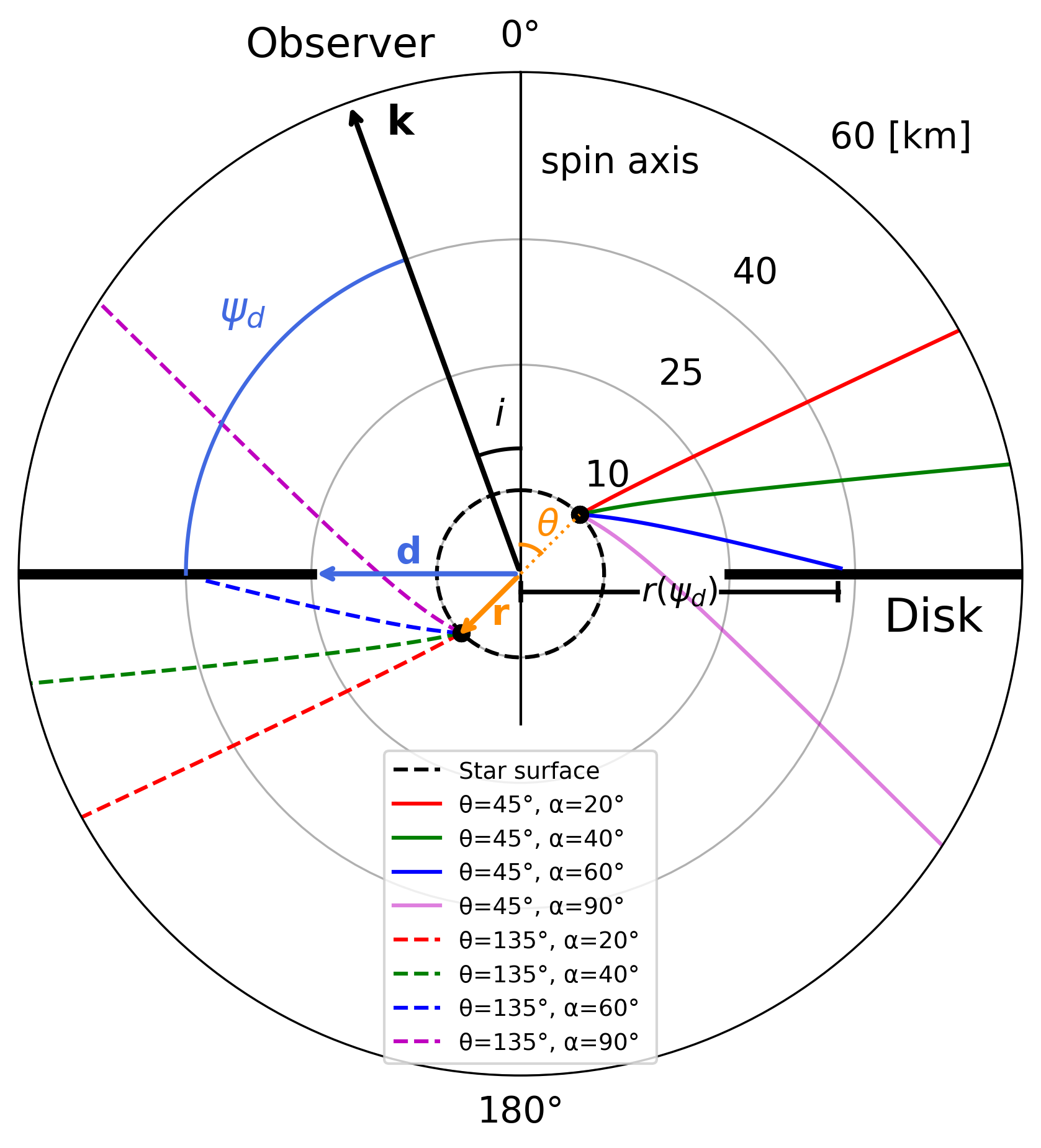}
    \caption{Schematic illustration of ray trajectories under disc occultation. Some rays pass through the region interior to the inner disc radius $R_{\rm in}$ ($\alpha = 90\degr$), while others are blocked by the disc ($\alpha = 60\degr$). The remaining rays propagate directly to infinity without intersecting the disc. All vectors indicate directions only; their lengths are used for visualization, while physically they are unit vectors. Note that this is a two-dimensional schematic: the observer, the spin axis, and the vector $\mathbfit{d}$ are projected onto the plane of the figure and may not be coplanar in three dimensions. Each ray corresponds to a specific value of $r(\psi_d)$; the annotation of $r(\psi_d)$ in the figure highlights only the case with $\alpha = 60\degr$.}
    \label{fig:Ray_trajectory}
\end{figure}

We use the approximate formulae given in \citet{Beloborodov2002} and \citet{Ibragimov2009} to determine whether a given light ray is blocked by the disc. In \citet{Beloborodov2002}, an approximate expression for the light bending of photons emitted from the NS surface was provided
\begin{equation}
r(\psi) =
\left[
\frac{r_{\mathrm{s}}^{2}\,(1-\cos\psi)^{2}}{4(1+\cos\psi)^{2}}
+ \frac{b^{2}}{\sin^{2}\psi}
\right]^{1/2}
- \frac{r_{\mathrm{s}}(1-\cos\psi)}{2(1+\cos\psi)} ,
    \label{eq:r_psi}
\end{equation}
where $r(\psi)$ gives the radial coordinate of the photon along its trajectory as a function the angular coordinate of the photon position $\psi$, with $\psi=0$ corresponding to the escape direction. Here, $r_{\rm s} = 2GM/c^2$ is the Schwarzschild radius, with $G$ the gravitational constant and $c$ the speed of light, and $b$ the impact parameter, given by
\begin{equation}
b = \frac{R_{\theta} \sin \alpha}{\sqrt{1-u}}, \quad u = \frac{r_{\rm s}}{R_{\theta}},
	\label{eq:impact factor}
\end{equation}
which represents the perpendicular distance from the star’s center to the photon’s trajectory as measured at infinity. $R_{\theta}$ is the local NS radius at the emission colatitude $\theta$ on an oblate stellar surface, and $\alpha$ is the emission angle with respect to the local outward radial vector. The photon trajectories emitted from the surface and propagating to infinity can be derived from this formula and are shown in Fig.~\ref{fig:Ray_trajectory}.

To compute the radius at which the light ray intersects the disc plane, $r(\psi_{\rm d})$, we use vector algebra to obtain the unit vector $\mathbfit{d}$ pointing to the intersection point. Because photon trajectories in Schwarzschild spacetime are planar, this vector lies along the intersection of the accretion disc plane with the plane defined by the line of sight $\mathbfit{k}$ and the radius-vector of the emission point $\mathbfit{r}$. We define the coordinate system with the $z$-axis along the disc axis and the line-of-sight lying in the $x$-$z$ plane. 
We then have 
\begin{equation}
\mathbfit{r} = (\sin\theta \cos\phi,\ \sin\theta \sin\phi,\ \cos\theta), \
\mathbfit{k} = (\sin i,\ 0,\ \cos i),
\end{equation}
where $i$ is the observer inclination, $\theta$ is the colatitude of the emission point and $\phi$ is the azimuthal angle. 
The normal to the trajectory plane defined by these two vectors is given by
\begin{equation}
\mathbfit{N} 
=(\sin\theta \sin\phi \cos i, \cos\theta \sin i - \sin\theta \cos\phi \cos i, -\sin\theta \sin\phi \sin i). 
\end{equation}
The direction vector $\mathbfit{d}$ along the line of intersection between this plane and the disc plane is given by
\begin{equation} 
\label{eq:dvector} 
\mathbfit{d}
=
\frac{
\left(
-\cos\theta \sin i + \cos i \sin\theta \cos\phi,\;
\sin\theta \cos i \sin\phi,\;
0
\right)
}{
\sqrt{\cos^2 i + \cos^2 \theta - 2 \cos i \cos\theta \cos\psi}
},
\end{equation}
where $\psi$ is the angle between $\mathbfit{k}$ and $\mathbfit{r}$ with 
\begin{equation}
\cos\psi = \mathbfit{k} \cdot \mathbfit{r} = \cos i \cos\theta + \sin i \sin\theta \cos\phi. 
\end{equation}
Equation \eqref{eq:dvector} follows \citet{Ibragimov2009} (their Eq.~C1), and is obtained by taking $\mathbfit{d} \propto \mathbfit{N} \times \mathbfit{z}$ 
and subsequently normalizing, where \mathbfit{z} = (0,0,1) is the disc normal. The angle $\psi_{\rm d}$ between this unit vector and the line of sight direction can then be calculated as
\begin{equation}
\cos\psi_{\rm d} = \mathbfit{d} \cdot \mathbfit{k}
= \frac{\cos i \cos\psi - \cos\theta}
{\sqrt{\cos^2 i + \cos^2 \theta - 2 \cos i \cos\theta \cos\psi}} .
	\label{eq:cos_psi_d}
\end{equation}
Substituting the resulting $\psi_{\rm d}$ into Eq.~\eqref{eq:r_psi} gives the radius at which the photon trajectory crosses the disc plane. If the intersection radius is smaller than the inner disc radius $R_{\rm in}$, the photon is assumed to pass through the region inside the disc inner edge without being blocked, as illustrated by the purple curve with $\alpha = 90\degr$ in Fig.~\ref{fig:Ray_trajectory}. If the photon intersects the disc plane at a radius larger than $R_{\rm in}$, the photon is considered to be occulted by the disc and therefore invisible. We assume that the disc extends to arbitrarily large radii, such that any intersection beyond $R_{\rm in}$ leads to occultation. This criterion has been implemented in the \acs{X-PSI} code as a new condition for determining photon visibility.\footnote{The \acs{X-PSI} code includes disc occultation from version v3.2, see \href{https://xpsi-group.github.io/xpsi}{https://xpsi-group.github.io/xpsi}.}

\subsection{Code verification}
To validate the reliability of our code, we performed a direct comparison with an independent code (hereafter IC) that includes disc obscuration, described in \citet{Poutanen2003}, \citet{Poutanen2006}, and \citet{Ibragimov2009}. The same code was also employed in \citet{Molkov2024} to fit the pulse profile of J1444.

In order to ensure a fully consistent comparison between the two implementations, we do not adopt the atmosphere model described in the previous sections. Instead, we used the same angular dependence for the radiation intensity as in \citet{Molkov2024}:
\begin{equation}
I(\mu) \propto 1 + h \cos \alpha',
    \label{eq:I_u}
\end{equation}
where $\alpha'$ is the emission angle relative to the local normal in the comoving reference frame. We adopt NS mass $1.4~{\rm M_\odot}$, radius $12~\mathrm{km}$, spin frequency $\nu = 447.9~\mathrm{Hz}$, and assume a spherical NS surface. 
The observer inclination is $i = 58\degr$, and we assume a primary circular hotspot centred at colatitude $\theta = 14\degr$, an inner disc radius of $R_{\rm in} = 24.5~\mathrm{km}$, and a phase shift of $0.45$. 
The secondary hotspot is assumed to be antipodal, and both hotspots have angular radius $\zeta = 33\degr$ and the anisotropy parameter $h = -0.65$. Because the physical assumptions and computational logic of \acs{X-PSI} and the code used in \citet{Molkov2024} are identical, the two implementations are expected to produce nearly indistinguishable results.\footnote{The qualifier ``nearly'' reflects possible small deviations arising from numerical and geometrical resolution effects.}

\begin{figure}
	\includegraphics[width=\columnwidth]{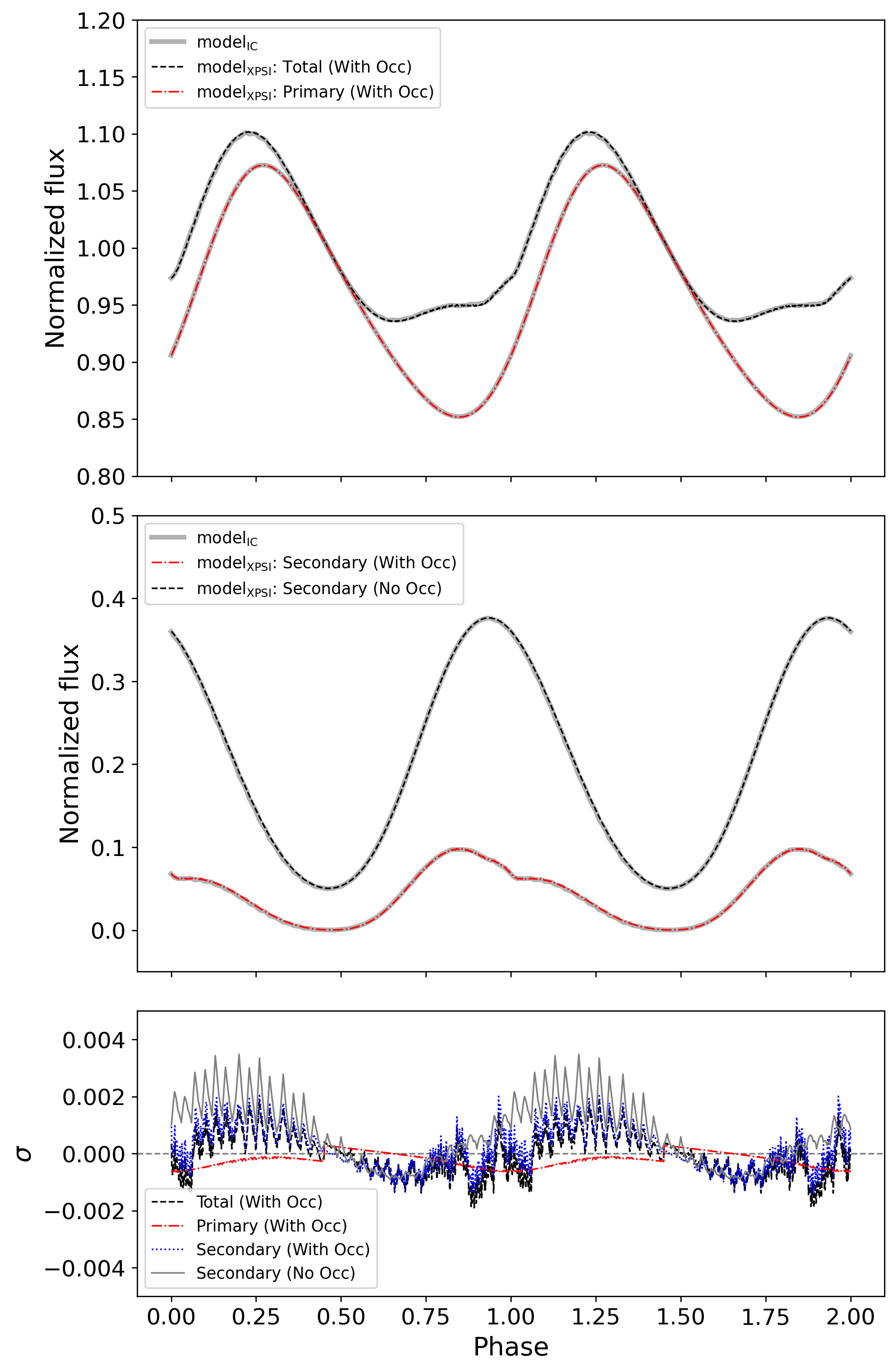}
    \caption{Normalized pulse profiles computed with \acs{X-PSI} and with the independent code used in \citet{Molkov2024}. The \acs{X-PSI} results are shown with coloured curves, while the independent code (IC) results are displayed as light gray dashed curves in the background. The top panel shows the total pulse profile including disc occultation, together with the contribution from the primary hotspot alone. The middle panel presents the emission from the secondary hotspot, computed both with and without disc occultation. The bottom panel shows the corresponding fractional differences, $\sigma$, for each of the four curves.}
    \label{fig:Cross_check}
\end{figure}

The comparison is shown in Fig.~\ref{fig:Cross_check}. The black dashed and red dotted-dash curves correspond to the results obtained with \acs{X-PSI}, while the light gray solid curve represents the results produced by the IC. The two results are in excellent agreement, with no visually discernible differences. We further quantified the discrepancy by computing
\begin{equation}
\sigma = \frac{\mathrm{model_{XPSI}} - \mathrm{model_{IC}}}{\mathrm{model_{IC}}},
    \label{eq:sigma}
\end{equation}
which remains below 0.4\% across all phases, as shown in the bottom panel. Such small differences are likely attributable to minor implementation details, such as the hotspot surface resolution or the angular resolution used in ray tracing. This is well below the typical Poisson noise level, indicating that our implementation of disc blocking is sufficiently accurate for our current studies \citep[for a related discussion see][]{Choudhury24b}.

\begin{figure*}
\includegraphics[width=1.9\columnwidth]{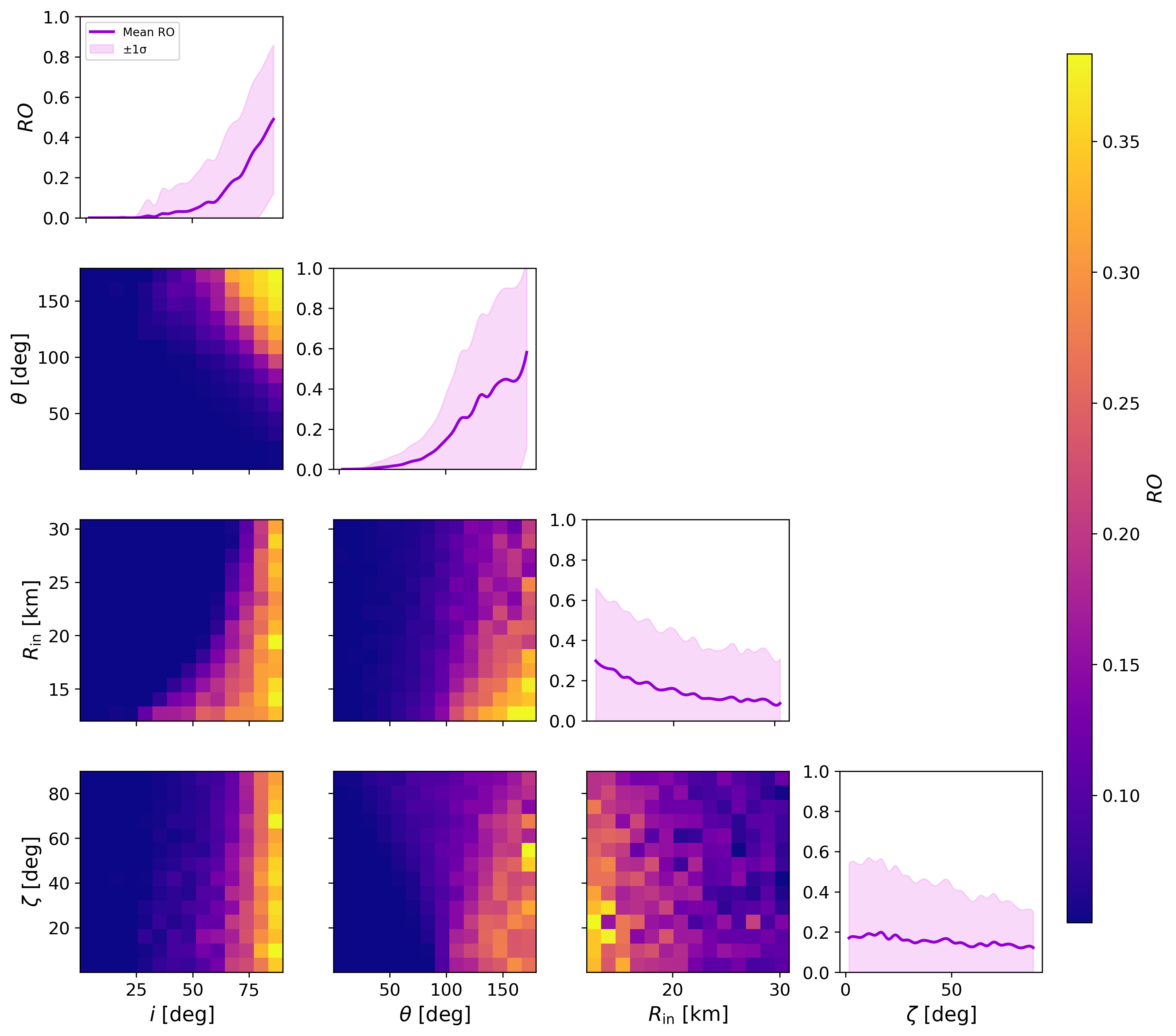}
\caption{Mean relative obscuration fraction as a function of the inclination angle $i$, the hotspot colatitude $\theta$, the inner disc radius $R_{\rm in}$, and the hotspot angular radius $\zeta$, based on $20000$ uniformly sampled points. The colour scale shows the relative obscuration fraction.
}
\label{fig:Relative_Obscuration}
\end{figure*}

\section{Effect of disc occultation across parameter space}\label{sec: Effect of disc occultation across parameter space}

In this section, we quantify the strength of disc occultation under different geometric conditions and identify the regions of parameter space in which this effect becomes significant. 
We select the four parameters most relevant to the strength of disc occultation: 
the inclination angle $0\degr<i<90\degr$, 
the hotspot colatitude $0\degr<\theta<180\degr$, 
the inner disc radius $12~{\rm km}<R_{\rm in}<30~{\rm km}$, 
and the hotspot angular radius $0\degr<\zeta<90\degr$. We adopt an oblate star with $M = 1.4~{\rm M}_{\odot}$ and $R_{\rm eq} = 12~{\rm km}$, with the latter also defining the lower bound of $R_{\rm in}$. To clearly isolate the effects of parameter variations, we consider only a single circular hotspot represented by these parameter combinations. We generate 20000 random combinations by uniformly sampling $\cos i\in(0,1)$, $\cos\theta\in(-1,1)$, $R_{\rm in}\in(12,30)~{\rm km}$, and $\zeta\in(0,90\degr)$, and compute the relative obscuration ($RO$) to quantify the degree of disc obscuration. We define $RO$ as
\begin{equation}
RO = \frac{F_{\rm no} - F_{\rm o}}{F_{\rm no}}.
\label{eq:RO}
\end{equation}
Here $F_{\rm o}$ and $F_{\rm no}$ are the total observed fluxes  with and without disc occultation. Samples with $F_{\rm no} = 0$ are excluded, as they correspond to emission fully obscured by the star itself and hence not subject to disc occultation. A larger $RO$ corresponds to a higher fraction of disc-eclipsed emission and therefore a stronger disc occultation effect. Fig.~\ref{fig:Relative_Obscuration} presents a multi-parameter projection plot showing the dependence of $RO$ on the four parameters. The off-diagonal panels represent the mean $RO$ in colour within grid cells defined by each pair of parameters, while the diagonal panels show how the mean $RO$ varies with individual parameters.

Fig.~\ref{fig:Relative_Obscuration}  reveals a strong sensitivity of disc occultation to the inclination angle $i$. As $i$ increases, the obscuration becomes progressively stronger, with a critical transition occurring at $i \simeq 50\degr$--$60\degr$. Beyond this range, the effect of disc occultation becomes important, with $RO$ rising rapidly as the inclination increases. This correlation is physically intuitive, since increasing the inclination angle brings the line of sight closer to the disc plane. In this geometry, visible photon trajectories intersect the disc plane at larger distances from the NS than in the low-$i$ case, making it less likely for photons to pass within the inner disc radius.

The dependence on the hotspot colatitude $\theta$ is asymmetric between the two hemispheres. Disc occultation primarily affects emission from the southern hemisphere, where it becomes weaker as the hotspot moves from the polar region toward the equator. Emission from the equatorial region intersects the disc plane at smaller radii, corresponding to smaller values of $r(\psi_{\rm d})$, and is therefore more likely to escape from within the inner disc radius without being blocked. In contrast, for hotspots in the northern hemisphere, obscuration is essentially absent although it may increase slightly as the hotspot approaches the equator. Surface elements with $\theta < 90\degr$ are always visible, such that obscuration in the northern hemisphere arises only when the hotspot extends across the equator (i.e., $\theta + \zeta > 90\degr$).

For the inner disc radius $R_{\rm in}$, a larger value naturally opens a wider region between the NS surface and the disc, allowing a greater fraction of photons to pass through and thus reducing the overall blocking fraction. However, since $i$ and $\theta$ induce much larger variations in $r(\psi_{\rm d})$, the impact of $R_{\rm in}$ is subdominant. Owing to the random sampling of the parameter space, the correlation between the blocking fraction and $R_{\rm in}$ remains relatively weak, exhibiting a mild negative trend.

The hotspot angular radius $\zeta$ shows no clear correlation with $RO$ within the explored parameter range. This indicates that the size of the emitting region has little impact on the relative disc occultation, and that the effect is governed by the global viewing geometry rather than the detailed extent of the hotspot. It appears that as the hotspot size increases, both the obscured and unobscured emission fractions increase in a similar manner, leaving the overall fractional obscuration largely unchanged.

Based on these results, we conclude that disc occultation must be treated with particular care when modelling systems viewed at moderate to high inclinations. For NSs with $i \gtrsim 50\degr$, disc occultation can substantially modify the pulse profiles and should not be neglected in parameter recovery using real observational data. The inclination angle, which may be independently constrained through polarization measurements, therefore plays a critical role in determining whether disc occultation needs to be included. In contrast, for systems viewed at low inclinations, the effect is likely to be much less important.

Based on the analysis in this section, we have quantified the degree of disc obscuration and demonstrated that, for certain regions of parameter space, disc blocking cannot be neglected in pulse profile modelling. The resulting differences may have some impact on the recovery of parameters. In the next section, we move on to the sampling stage and examine how disc blocking influences the inferred parameters. If neglecting disc blocking leads to biased or incorrect inferences, this would provide strong justification for adopting models that explicitly include disc blocking when analysing observational data.

\begin{table*}
\centering
\small
\setlength{\tabcolsep}{3.5pt}
\caption{Summary of sampling runs and model configurations used in the parameter recovery analysis in Scenarios A and B. For a description of the columns see the text.}
\label{tab:sampling_runs_AB}

\begin{tabular}{@{}c c c c c c c c c c c c c@{}}
\hline
Scenario & Run & Fix $i$ & Fix $\theta_{\rm p}$ & Disc Emis$^{\rm a}$ & Occ (model)$^{\rm b}$ & Occ (data)$^{\rm c}$ & Multimode & Antipode
& Live points & $N_{68\%}^{\rm cov}$$^{\rm d}$ & $N_{90\%}^{\rm cov}$$^{\rm e}$ & $N_{\rm free}$$^{\rm f}$ \\
\hline
A & 1 & \checkmark & \checkmark & $\times$ & $\times$  & $\times$  & $\times$ & \checkmark & 1000  & 8 & 13 & 13 \\
A & 2 & \checkmark & \checkmark & $\times$ & $\times$  & $\times$  & \checkmark & $\times$ & 1000  & 5/7 $^{\rm g}$ & 8/11 &  15 \\
A & 3 & \checkmark & \checkmark & \checkmark & $\times$ & \checkmark & $\times$ & $\times$ & 1000 & 10 & 13 & 17 \\
A & 4 & \checkmark & \checkmark & \checkmark & \checkmark & \checkmark & $\times$ & $\times$ & 1000 & 10 & 12 & 17 \\
A & 5 & \checkmark & $\times$ & \checkmark & $\times$ & \checkmark & $\times$ & $\times$ & 1000 & 7 & 9 & 18 \\
A & 6 & \checkmark & $\times$ & \checkmark & \checkmark & \checkmark & $\times$ & $\times$ & 1000 & 5 & 8 & 18 \\
A & 7 & $\times$ & $\times$ & \checkmark & \checkmark & \checkmark & \checkmark & $\times$ & 2000 & 6 & 7 & 19 \\
\hline
B & 8 & \checkmark & \checkmark & $\times$  & $\times$  & $\times$  & $\times$ & $\times$ & 1000  & 15 & 15 & 15 \\
B & 9 & \checkmark & \checkmark &  \checkmark & $\times$ & \checkmark & $\times$ & $\times$ & 1000 & 15 & 15 & 17 \\
B & 10 & \checkmark & \checkmark &  \checkmark & \checkmark & \checkmark & $\times$ & $\times$ & 1000 & 15 & 16 & 17 \\
B & 11 & \checkmark & $\times$ &  \checkmark & $\times$ & \checkmark & $\times$ & $\times$ & 1000 & --$^{\rm h}$ & -- &18 \\
B & 12 & \checkmark & $\times$ &  \checkmark & $\times$ & \checkmark & \checkmark & $\times$ & 2500 & 1/14 & 1/16 &18 \\
B & 13 & \checkmark & $\times$ &  \checkmark & \checkmark & \checkmark & \checkmark & $\times$ & 1000 & 13 & 16 & 18 \\
B & 14 & $\times$ & $\times$ &  \checkmark & $\times$ & \checkmark & $\times$ & $\times$ & 1000 & 15 & 17 & 19 \\
B & 15 & $\times$ & $\times$ &  \checkmark & \checkmark & \checkmark & $\times$ & $\times$ & 1000 & 16 & 17 & 19 \\
\hline
\end{tabular}

\vspace{2pt}
\parbox{\textwidth}{\footnotesize
\textbf{}
$^{\rm a}$ Disc emission included in model and synthetic data.

$^{\rm b}$ Disc occultation included in the model.

$^{\rm c}$ Disc occultation included in the synthetic data.

$^{\rm d}$ Number of parameters whose true values lie within the 68\% credible interval.

$^{\rm e}$ Number of parameters whose true values lie within the 90\% credible interval.

$^{\rm f}$ Number of free parameters in each run.

$^{\rm g}$ Two values are reported because two posterior modes are identified; the left and right values correspond to mode~1 and mode~2, respectively.

$^{\rm h}$ The results under this configuration are seed-dependent; see the main text for further details.
}
\end{table*}

\section{Parameter inference with disc occultation}\label{sec:Parameter inference with disc occultation}
In this section, we perform Bayesian parameter inference to assess the impact of disc occultation on the recovered model parameters. We analyse synthetic NICER data using models both including and neglecting disc occultation, and compare the resulting posterior distributions to quantify the biases and uncertainties introduced by disc occultation.

In Section \ref{sec:parameter inference}, we first describe the Bayesian framework adopted for parameter inference. Section \ref{sec:parameter set} presents two parameter configurations (Scenarios A and B) and their associated priors. Owing to the intrinsic complexity of the \stu{} two hotspot model, we adopt a staged approach in our parameter recovery. For each scenario, we first consider a simplified case without a disc, in which neither disc emission nor disc occultation is included, and then progressively extend the analysis to models that include a disc. The results of these analyses are presented in Section~\ref{sec:sampling result}.

\subsection{Parameter Inference}\label{sec:parameter inference}

We adopt a Bayesian framework for parameter inference and employ a nested sampling strategy to explore the resulting posterior distributions. Our goal is to quantify how well the model parameters can be recovered from data under different modelling assumptions, and to assess the impact of disc occultation on the inferred results. The posterior probability distribution of the model parameters $\theta$, given a dataset $D$ and a model $M$, is
\begin{equation}
P(\theta \mid D, M) = \frac{P(D \mid \theta, M)\, P(\theta \mid M)}{P(D \mid M)} .
    \label{eq:Bayesian}
\end{equation}
Here, $P(D \mid \theta, M)$ is the likelihood function, $P(\theta \mid M)$ denotes the prior distribution, and $P(D \mid M)$ is the Bayesian evidence. In this analysis, the dataset $D$ consists of synthetic NICER pulse profiles generated using different sets of input parameters. The likelihood function is constructed assuming Poisson statistics for the NICER count data. We focus on the posterior distributions, and by comparing them with the input parameters, we quantify how accurately the model parameters are constrained by the data.

Sampling of the posterior distributions is carried out using a nested sampling approach implemented with \multinest{} \citep{Skilling2004,Feroz2009,Buchner2016,Feroz2019}, which is well suited for exploring high-dimensional parameter spaces and identifying multiple disconnected regions of high likelihood. In particular, the use of multimodal sampling allows the algorithm to identify distinct parameter vectors that provide comparably good fits to the same dataset. This capability is essential for our analysis, as degeneracies between model parameters can lead to multiple modes in the posterior distribution, especially for the intrinsically complex \stu{} two hotspot model.

We perform a series of sampling runs with varying configurations, ranging from simplified setups without a disc to more complex models including disc emission and occultation. In some runs, parameters such as the inclination angle or hotspot colatitude are fixed, to break degeneracies. This choice is based on the fact that polarization measurements may provide strong independent constraints on these parameters in realistic observational scenarios. An overview of the sampling configurations adopted in this work is provided in Table~\ref{tab:sampling_runs_AB}. We discuss the contents of the table in detail in the subsequent sections.

\begin{table*}
\caption{\label{tab:scenario_parameters}
Model parameters for Scenarios A and B. For each parameter, the adopted value(s) in the corresponding scenario and the prior distribution used in the Bayesian analysis are listed. A dash (--) indicates that the parameter is the same as in Scenario A.}
\centering
\begin{tabular}{llll}
\hline
Parameter (unit) & Scenario A & Scenario B & Prior distribution \\ 
\hline
\multicolumn{4}{c}{Pulsar} \\
$M$ ($\rm M_\odot$) & 1.4 & -- & $U(1.0,\,3.0)$$^{\rm a}$ \\
$R_{\rm eq}$ (km) & 12 & -- & $U(3R_{\rm g}(1),\,16)$ $^{\rm b}$ \\
$D$ (kpc) & 8.0 & -- & $N(8.0,\, 1.0)$ \\
$\cos i$ (-) & 0.53 & -- & $U(0,\,1)$ \\
$f$ (Hz) & 447.9 & -- & fixed \\ 
\hline
\multicolumn{4}{c}{Primary hot spot} \\
$\phi_{\rm p}$ (-) & 0 & - & $U(0,\,1)$ \\
$\cos\theta_{\rm p}$ (-) & 0.97 & 0.5 & $U(-1,\,1)$$^{\rm c}$ \\
$\zeta_{\rm p}$ (deg) & 33 & -- & $U(0.001,\,90)$ \\
$T_{\rm bb,p}$ (keV) & 1 & -- & $U(0.5,\,1.5)$ \\
$T_{\rm e,p}$ (keV) & 50 & -- & $U(20,\,100)$ \\
$\tau_{\rm p}$ (-) & 1 & -- & $U(0.5,\,3.5)$ \\
\hline
\multicolumn{4}{c}{Secondary hot spot} \\
$\phi_{\rm s}$ (-) & 0 & - & $U(0,\,1)$ \\
$\cos \theta_{\rm s}$ (-) & $-$0.97 & $-$0.5 & $U(-1,\,1) $ \\
$\zeta_{\rm s}$ (deg) & 33 & -- & $U(0.001,\,90)$ \\
$T_{\rm bb,s}$ (keV) & 1 & -- & $U(0.5,\,1.5)$ \\
$T_{\rm e,s}$ (keV) & 50 & -- & $U(20,\,100)$ \\
$\tau_{\rm s}$ (-) & 1 & -- & $U(0.5,\,3.5)$ \\
\hline
\multicolumn{4}{c}{Disc} \\
$T_{\rm in}$ (keV) & 0.37 & -- & $U(0.01,\,0.6)$ \\
$R_{\rm in}$ (km) & 24.5 & -- & $U(R,\,R_{\rm co})$$^{\rm d~e}$  \\
\hline
\multicolumn{4}{c}{Absorption} \\
$N_{\rm H}$ ($10^{21}\,\mathrm{cm^{-2}}$) & 29 & -- & $N(29.0,\, 4.0)$  \\
\hline
\end{tabular}

\vspace{2pt}
\parbox{\textwidth}{\footnotesize
\textbf{}

$^{\rm a}$ In addition to the explicit bounds, the prior includes hard physical cuts on compactness.

$^{\rm b}$ $R_{\rm g}(1) = GM/c^2$ is the gravitational radius for $M = 1~{\rm M}_\odot$. 

$^{\rm c}$ Additional rejection conditions enforce ordering and non-overlap of the two hot spots.

$^{\rm d}$ Here $R_{\rm co}$ is the co-rotation radius, $R_{\rm co} = ({GM}/{4f^2 \pi^2})^{1/3}$.

$^{\rm e}$ The priors on $M$, $R_{\rm eq}$, and $R_{\rm in}$ are not independent, as samples are rejected when $R_{\rm in} > R_{\rm co}$ or $R_{\rm in} < R$.
}
\end{table*}

\begin{table*}
\caption{\label{tab:parameter CI}
Credible intervals within which the true value of each parameter is recovered, for all runs listed in Table \ref{tab:sampling_runs_AB}. For each parameter, the reported value denotes the smallest credible interval (68\%, 90\%, or 95\%) that contains the true value, or indicates that the true value lies outside the 95\% credible interval. Run 11 is not shown as its results are seed-dependent.}
\centering
\small
\begin{tabular}{|*{20}{c|}}
\hline
Run & $M$ & $R_{\rm eq}$ & $D$ & $\cos i$ & $\phi_{\rm p}$ & $\cos \theta_{\rm p}$ & $\zeta_{\rm p}$ & $T_{\rm bb,p}$ & $T_{\rm e,p}$ & $\tau_{\rm p}$ & $\phi_{\rm s}$ & $\cos \theta_{\rm s}$ & $\zeta_{\rm s}$ & $T_{\rm bb,s}$ & $T_{\rm e,s}$ & $\tau_{\rm s}$ &  $R_{\rm in}$ & $T_{\rm in}$ & $N_{\rm H}$ \\
\hline
1 & 68 & 68 & 68 & -- & 68 & -- & 90 & 90 & 90 & 90 & -- & -- & 68 & 68 & 68 & 68 & -- & -- & 90 \\ 
2 (mode 1) & 90 & 68 & 68 & -- & 68 & -- & 68 &90 & 95 & >95 & >95 & >95 & >95 & >95 & 68 & >95 & -- & -- & 90  \\ 
2 (mode 2) & 68 & 90 & 68 & -- & 90 & -- & >95 & 68 & >95 & >95 & 68 & 90 & 68 & 68 & 68 & 90 & -- & -- & 95  \\ 
3 & 90 & 68 & 68 & -- & 68 & -- & 68 & 95 & 68 & 68 & 90 & >95 & 95 & 68 & 68 & 95 & 68 & 68 & 90 \\ 
4 & 95 & 68 & 68 & -- & 68 & -- & 68 & >95 & 68 & 68 & 68 & >95 & >95 & 68 & 68 & >95 & 90 & 68 & 90  \\ 
5 & 95 & >95 & 68 & -- & 90 & >95 & >95 & >95 & 68 & 68 & 68 & >95 & >95 & 95 & 68 & 90 & 68 & >95 & 68 \\ 
6 & >95 & >95 & 68 & -- & 90 & >95 & >95 & >95 & 68 & 90 & 68 & >95 & >95 & 68 & 68 & 95 & 90 & >95 & >95 \\ 
7 & >95 & >95 & 68 & >95 & 90 & >95 & >95 & >95 & 68 & 68 & >95 & >95 & >95 & >95 & 68 & 68 & >95 & >95 & 68 \\ 
\hline
8 & 68 & 68 & 68 & -- & 68 & -- & 68 & 68 & 68 & 68 & 68 & 68 & 68 & 68 & 68 & 68 & -- & -- & 68 \\ 
9 & 68 & 68 & 68 & -- & 68 & -- & 68 & 68 & 68 & 68 & 68 & 68 & >95 & 68 & 68 & 68 & 68 & 68 & 95 \\ 
10 & 68 & 68 & 68 & -- & 68 & -- & 68 & 68 & 68 & 68 & 68 & 68 & 90 & 68 & 68 & 68 & 68 & 68 & >95 \\ 
12 (mode 1) & >95 & >95 & >95 & -- & >95 & >95 & >95& >95 & >95 & >95 & >95 & >95 & >95 & >95 & >95 & >95 & 68 & >95 & >95 \\ 
12 (mode 2) & 68 & 68 & 68 & -- & 68 & 90 & 68 & 68 & 68 & 68 & 68 & 90 & >95 & 68 & 68 & 68 & 68 & 68 & 95\\
13 & 68 & 68 & 68 & -- & 68 & 90 & 68 & 68 & 68 & 68 & 68 & 90 & 95 & 68 & 68 & 68 & 68 & 90 & >95 \\
14 & 68 & 68 & 68 & 68 & 68 & 90 & 68 & 68 & 68 & 68 & 68 & 68 & >95 & 90 & 68 & 68 & 68 & 68 & 95\\
15 & 68 & 68 & 68 & 68 & 68 & 90 & 68 & 68 & 68 & 68 & 68 & 68 & 95 & 68 & 68 & 68 & 68 & 68 & >95\\

\hline
\end{tabular}
\end{table*}

\subsection{Parameter configurations and priors} \label{sec:parameter set}
We consider two representative parameter configurations (Scenarios A and B) to generate synthetic NICER datasets for subsequent parameter inference. The two scenarios share the same underlying stellar, atmosphere, and disc parameters, differing only in hotspot colatitudes. This setup enables a controlled comparison of the impact of hotspot location on sampling performance. All synthetic datasets assume an exposure time of 132 ks.

\subsubsection{Scenario A: Best fit parameters of J1444} \label{sec:parameter of scenario A}
Scenario A adopts the set of parameters reported by \citet{Molkov2024}; their best-fit description of the NICER pulse profile of J1444, which can be described as a parameter vector within our \stu{} model. As emphasized in their work, this parameter set is not unique in reproducing the observed pulse profile. However, it corresponds to a configuration in which disc occultation plays a relatively prominent role in shaping the pulse profile due to the large $i$ and $\theta_{\rm s}$, as illustrated in Fig.~\ref{fig:Relative_Obscuration}.

The input parameters and prior distributions adopted for Scenario A are listed in Table \ref{tab:scenario_parameters}. This scenario shows an antipodal hotspot configuration, in which both hotspots are located close to the spin poles. If the hotspots of AMPs are formed by accreting material channelled onto the NS surface, this configuration naturally corresponds to a small magnetic inclination angle, i.e., a small angle between the magnetic and spin axes. Fig.~\ref{fig:Geometry} shows a schematic illustration of the geometry and the corresponding bolometric pulse profiles computed with and without disc occultation, exhibiting a clear difference near the flux minimum.

We adopt a NS mass of $M = 1.4\,{\rm M_\odot}$ with a uniform prior $U(1,\,3)\,{\rm  M_\odot}$, and a radius of $R_{\rm eq} = 12~{\rm km}$ with a uniform prior $U(3R_{\rm g}(1),\,16~{\rm km})$, where $R_{\rm g}(M/{\rm  M}_{\odot}) = (M/{\rm M}_\odot)G{\rm M}_{\odot}/c^2$ is the gravitational radius. In addition to these explicit bounds, we impose a compactness constraint ${R_{\rm pole}}/{R_{\rm g}(M/{\rm M}_{\odot})} > 2.9$, where $R_{\rm pole}$ is the polar radius \citep[see e.g.][]{Gandolfi2012}. The source distance follows a Gaussian
prior with $8.0 \pm 1.0$~kpc, truncated at $\pm 5\sigma$. The inclination angle satisfies $i < 90\degr$,
implemented through a uniform prior on $\cos i \in (0,1)$. The spin
frequency is fixed at $f = 447.9$~Hz, corresponding to the frequency of J1444.

We assume two hotspots with centre colatitudes $\theta_{\rm p} = 14\degr$ and $\theta_{\rm s} = 166\degr$. The priors are uniform in $\cos\theta$ ($\cos\theta \sim U(-1,1)$), corresponding to an isotropic surface distribution, with the ordering constraint $\theta_{\rm p} < \theta_{\rm s}$ imposed. The angular radii of both hotspots are $33\degr$ and drawn from a uniform prior $U(0.001\degr,\,90\degr)$. Hotspot configurations that lead to overlapping emitting regions are rejected. The three atmospheric parameters are assumed to be identical for both hotspots, with values listed in Table~\ref{tab:scenario_parameters}.

For the accretion disc, we adopt an inner disc temperature of $T_{\rm in} = 0.37$~keV and assign a uniform prior of $U(0.01,0.6)$~keV. Under the standard thin-disc framework \citep{Shakura1973}, this corresponds to an accretion rate of $\dot{M} \sim 10^{-10}~M_\odot~\mathrm{yr}^{-1}$, typical for low-mass X-ray binaries \citep[e.g.][]{Galloway2008}. 
The inner disc radius is $R_{\rm in} = 24.5$~km, and is expected to lie between the NS radius and the corotation radius given by Eq.~\eqref{eq:rco}.
We enforce $R_{\rm eq} < R_{\rm in} < R_{\rm co}$ by rejecting samples outside this range, ensuring the system remains in the accreting regime. The interstellar absorption column density is centred at $N_{\rm H} = 29~\times~10^{21}\,{\rm cm^{-2}}$, following \citet{Ng2024}, and
is assigned a Gaussian prior with $\mu = 29$ and $\sigma = 4$, cut-off at $\pm 5\sigma$.

\subsubsection{Scenario B: J1444-base parameters with equatorial hotspots} \label{sec:parameter of scenario B}
\begin{figure}
	\includegraphics[width=\columnwidth]{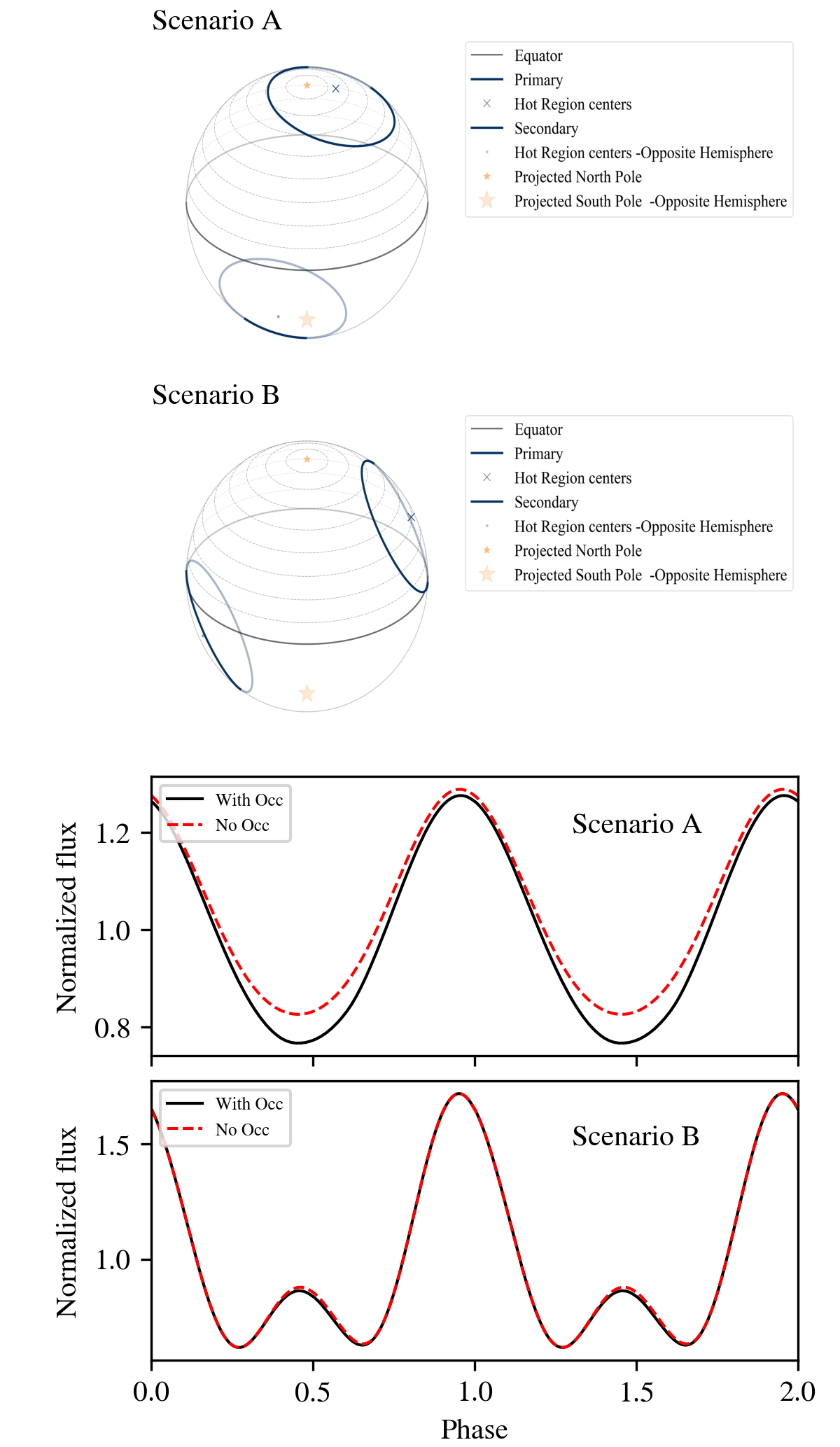}
    \caption{Schematic illustrations of the hotspot geometries and the corresponding pulse profiles for Scenario A and Scenario B. The first and second rows show the hotspot geometries for Scenario A and Scenario B. In both cases, the hotspots are antipodal, lying close to the poles in Scenario A and near the equator in Scenario B. The line-of-sight vector is set to $(0, 1, 0.5)$ for clarity, with the $z$-axis aligned with the spin axis. The third and fourth rows show the corresponding normalized bolometric pulse profiles for two scenarios. The black solid and red dashed curves represent the cases with and without disc occultation, illustrating the impact of disc blocking.}
    \label{fig:Geometry}
\end{figure}

Scenario B is derived from Scenario A by modifying the hotspot colatitudes, while keeping all other parameters unchanged. We reposition the hotspot colatitudes to $\theta_{\rm p} = 60\degr$ and $\theta_{\rm s} = 120\degr$. This configuration is chosen for two reasons. First, under this geometry, variations in the visibility of the two hotspots lead to pulse profile features that differ significantly from those in Scenario A, potentially affecting parameter recovery. Second, Scenario A places the secondary hotspot colatitude close to the boundary of its prior distribution, and parameters near prior boundaries may hinder accurate recovery of the true solution. In Scenario B, both hotspots are located closer to the stellar equator and well within the interior of the prior. The geometry and pulse profile of this configuration are shown in Fig.~\ref{fig:Geometry}, a slight difference between the profiles with and without occultation can be seen around phase $\sim 0.4$. Except for these two parameters, all other parameter values and prior distributions are identical to those in Scenario A.

\subsection{Parameter recovery results} \label{sec:sampling result}
In this section, we present Bayesian inference results for the synthetic datasets. For each scenario, we perform a series of sampling runs and compare the inferred posteriors with the true parameters to quantify biases in recovery.

\subsubsection{Inference results in Scenario A} \label{sec:inference of scenario A}

Table \ref{tab:sampling_runs_AB} summarises the parameter recovery runs for Scenario A under different configurations. Each column represents a specific setting of the sampling run. A check mark indicates that the condition is applied, while a cross means the opposite.

The columns Fix $i$ and Fix $\theta_{\rm p}$ denote runs in which the inclination angle $i$ and the primary hotspot colatitude $\theta_{\rm p}$ are fixed or free during sampling. In real observations, polarization measurements may constrain these parameters. The Disc Emis column shows whether disc emission is included in both the synthetic dataset and the sampling model. Occ (model) and Occ (data) specify disc occultation separately for the model ($M$) and dataset ($D$), as defined by Eq.~\eqref{eq:Bayesian}. The Multimode option, described in Section \ref{sec:parameter inference}, allows the sampler to identify multiple posterior modes for the same dataset. The Antipode column indicates whether the two hotspots are constrained to be antipodal during sampling (i.e., $\theta_{\rm s} = 180\degr - \theta_{\rm p}$ and $\phi_{\rm s} - \phi_{\rm p} = 0.5$). Live points gives the number of live points used in the nested sampling. We find that 1000 live points are sufficient in most cases, consistent with previous studies \citep{Dorsman2025}, and supported by tests with larger numbers of live points that yield very similar full posteriors. This is considered adequate for sampling the synthetic data used here, with a larger number of live points potentially required to fully ensure robust convergence or for applications to real data. A fixed sampling efficiency of 0.1 is used in all nested sampling runs. $N_{68\%}^{\rm cov}$ and $N_{90\%}^{\rm cov}$ denote the number of parameters for which the true values lie within the 68\% and 90\% credible intervals (CIs), respectively. $N_{\rm free}$ denotes the number of free parameters in each run. In the following sections, we refer to individual runs as Run followed by the corresponding row number in Table~\ref{tab:sampling_runs_AB}. The quantities $N_{68\%}^{\rm cov}$ and $N_{90\%}^{\rm cov}$ provide useful metrics for assessing parameter recovery. Table~\ref{tab:parameter CI} lists the smallest CI containing the true value for each parameter (68\%, 90\%, 95\%, or outside 95\%).

\begin{figure*}
	\includegraphics[width=2\columnwidth]{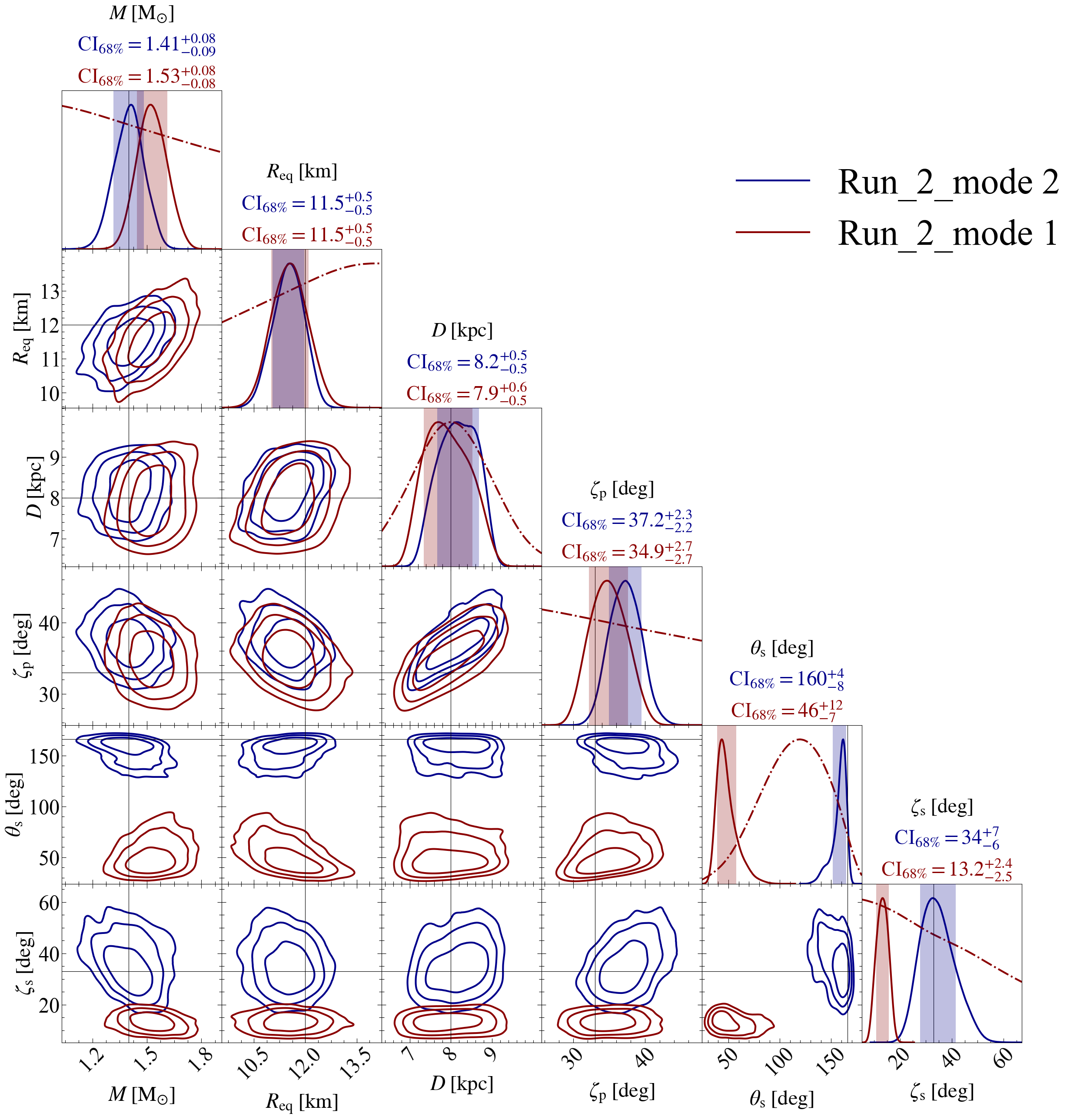}
    \caption{Posterior distributions for Run 2 listed in Table~\ref{tab:sampling_runs_AB}, showing only the six key parameters. The blue and red colours correspond to the two distinct solutions identified by the multimodal sampling. In the diagonal panels, the solid curves represent the one-dimensional marginalised posterior distributions, with the shaded regions indicating the 68 per cent credible intervals, while the dashed curves show the prior distributions. The off-diagonal panels display the two-dimensional posterior distributions, with contours enclosing the 68, 95, and 99.7 per cent credible regions. The black cross lines denote the true input parameter values. Note that the two modes are shown without weighting by their relative evidence.}
    \label{fig:run_2_corner}
\end{figure*}

As seen in Tables \ref{tab:sampling_runs_AB} and \ref{tab:parameter CI}, Scenario A represents a relatively challenging case for accurate parameter recovery. For Run~1, which has the fewest free parameters, the recovery performance is good: both $M$ and $R$ lie within the 68\% CI and all parameters fall within the 90\% CI. In Run~2, relaxing the antipodal symmetry constraint introduces additional freedom and leads to a more complex posterior. The multimodal sampling identifies two posterior modes with comparable statistical weight, as illustrated by the posterior distribution  in Fig.~\ref{fig:run_2_corner}. These two modes correspond to different parameter combinations. The difference in local log-evidence between modes~1 and~2 is 0.712, indicating that they cannot be statistically distinguished \citep{Kass1995}. Table~\ref{tab:parameter CI} shows that mode~2 lies closer to the input parameters, with improved recovery of the NS mass $M$ and radius $R_{\rm eq}$, and correctly identifies the secondary hotspot in the southern hemisphere. Similar bimodal posterior structures have also been reported in previous sampling studies (e.g., \citealt[Figs.~3.5 and~3.6]{Riley2019PhD}).

\begin{figure*}
	\includegraphics[width=2\columnwidth]{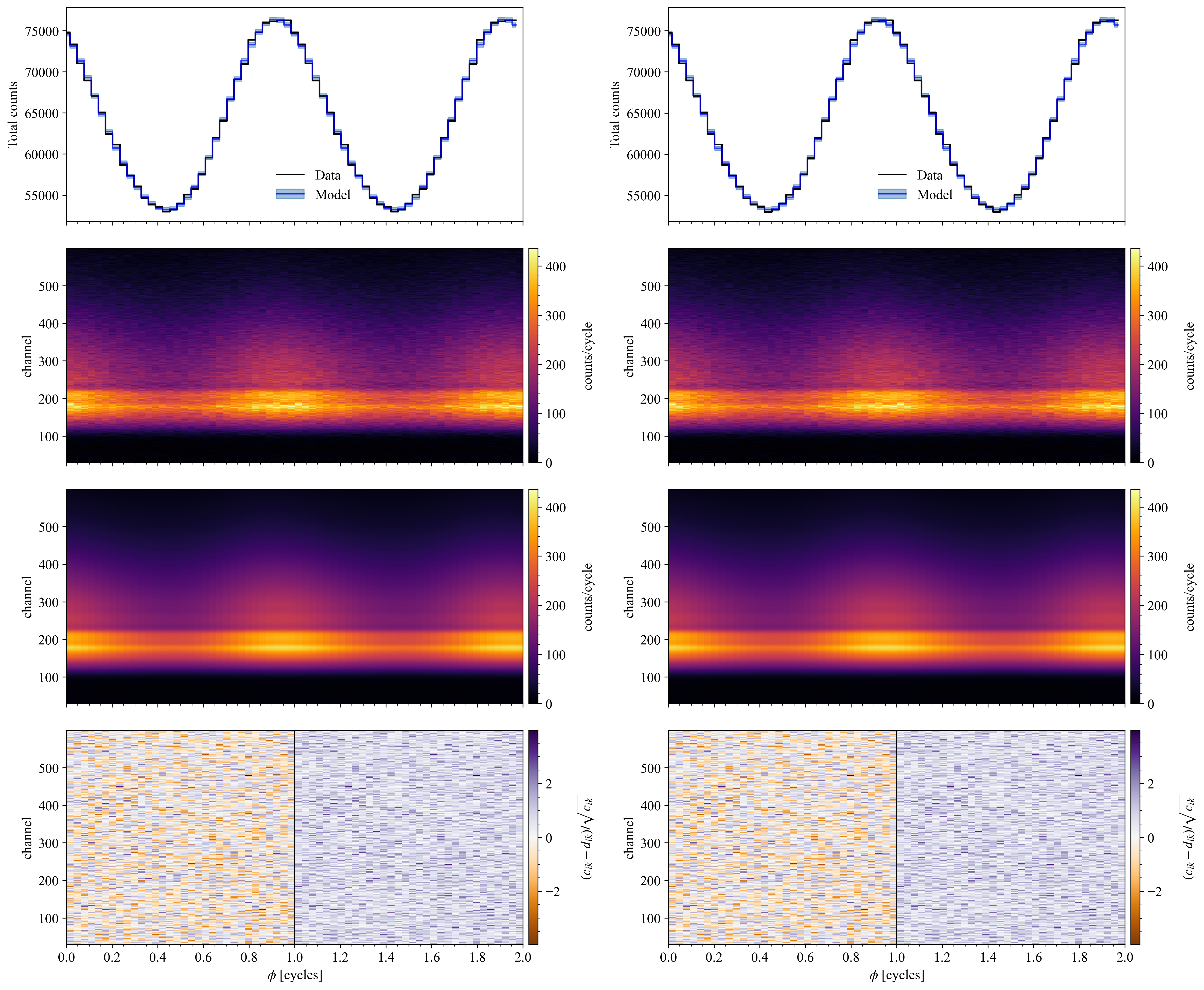}
    \caption{Comparison of the synthetic data, posterior-expected pulse profiles, and the corresponding residuals for Scenario A. Run 3 and Run 4 are represented in the left and right columns, respectively. The top panels show the bolometric pulse profiles of the data and the posterior-expected model. The black curves represent the synthetic data, while the blue shaded regions correspond to the 16–84 per cent credible intervals of the posterior distribution. 
    The second and third rows present the synthetic data and the posterior model predictions, with the former including a Poisson realisation and the latter corresponding to the underlying model expectation. Both are shown with a colour scale indicating the total NICER counts per cycle in each phase–energy bin for a total exposure of 132 ks. The fourth row shows the normalized residuals, computed as $(c_{\rm ik} - d_{\rm ik}) / \sqrt{c_{\rm ik}}$, where $c_{\rm ik}$ denotes the posterior expected pulse profile and $d_{\rm ik}$ the data. Here, $i$ indexes the energy channels and $k$ indexes the phase bins.
}
    \label{fig:run_3_4_residual}
\end{figure*}

For Runs~3 and 4, the synthetic data are generated with disc occultation included, while the sampling models either include or neglect this effect. The left and right columns in Fig.~\ref{fig:run_3_4_residual} correspond to Runs~3 and 4, respectively. The top row compares the synthetic bolometric data with the posterior-expected pulse profiles, demonstrating that both models closely reproduce the simulated data. The residuals (fourth row), computed from the data (second row) and the posterior-expected profiles (third row), show no significant structure and are consistent with Poisson noise, indicating statistically acceptable fits.

\begin{figure*}
	\includegraphics[width=2\columnwidth]{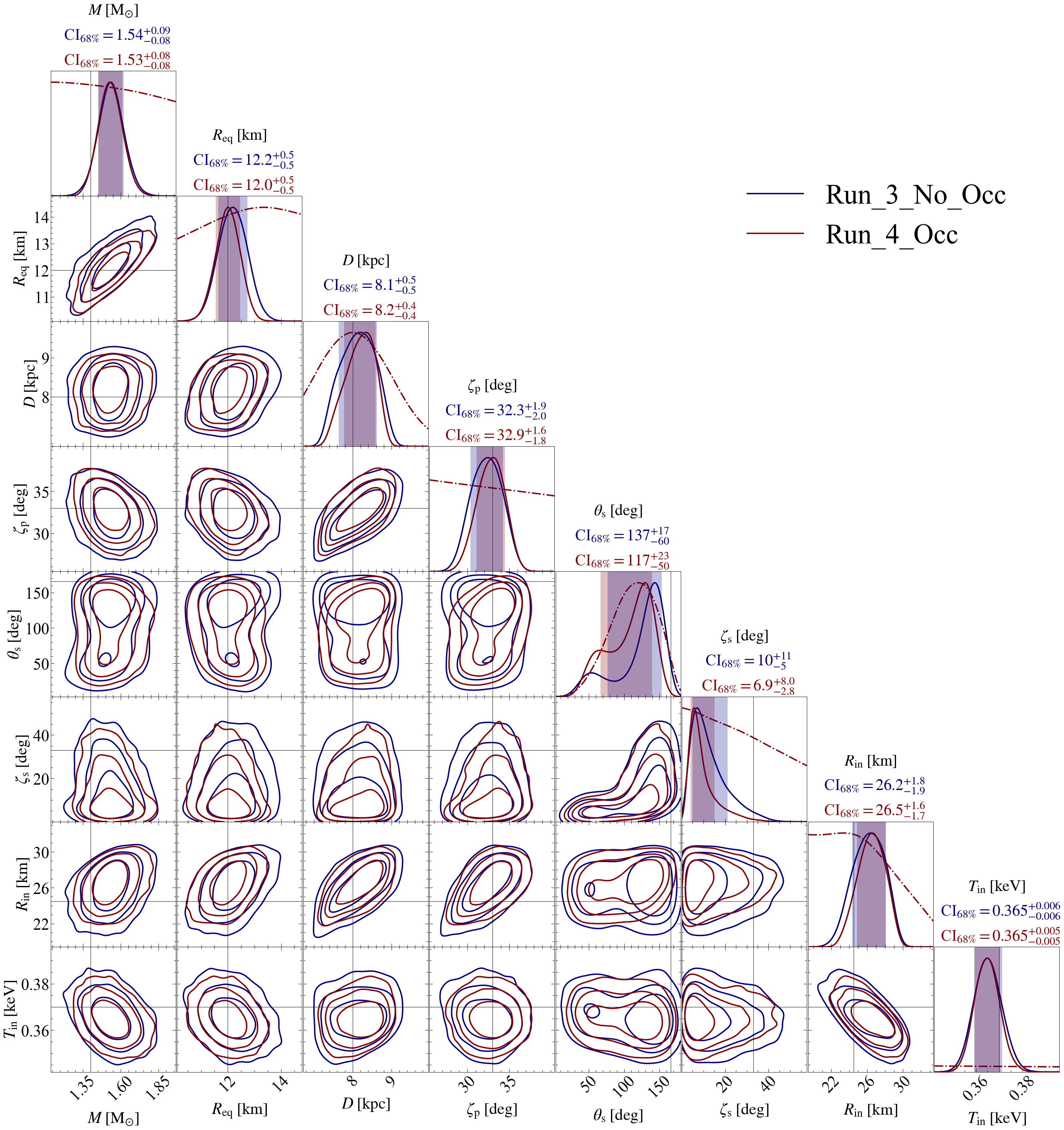}
    \caption{Posterior distributions for Runs 3 and 4 listed in Table~\ref{tab:sampling_runs_AB}, showing the eight key parameters. The definitions of all lines and shaded regions are the same as in Fig.~\ref{fig:run_2_corner}. The blue and red colours correspond to the sampling results obtained without and with disc occultation in the model, respectively (the data sets are created with disc occultation included).
}
    \label{fig:run_3_4_corner}
\end{figure*}

The posterior distributions for Runs~3 and 4 are shown in Fig.~\ref{fig:run_3_4_corner}. Although both models provide acceptable fits, the inferred posteriors show noticeable biases relative to the input parameters. In both runs, the NS mass falls outside the 68\% CI, and neither model is able to reliably constrain the location or angular size of the secondary hotspot. These biases persist in subsequent configurations with increasing model complexity.

To assess the impact of increased model complexity, Runs~5 and 6 allow $\theta_{\rm p}$ to vary freely, while Run~7 additionally frees $i$, resulting in all 19 parameters being sampled. With increasing freedom, the number of well-constrained parameters decreases, and key quantities, including $M$, $R_{\rm eq}$, and the hotspot geometry, become poorly recovered with the input parameters often lying outside the 95\% CI. These results suggest that Scenario~A corresponds to a regime in which reliable parameter recovery is difficult to achieve. The possible causes of this behaviour will be discussed in Section~\ref{sec:Geometric Dependence of Parameter Recovery}.

\begin{figure*}
	\includegraphics[width=2\columnwidth]{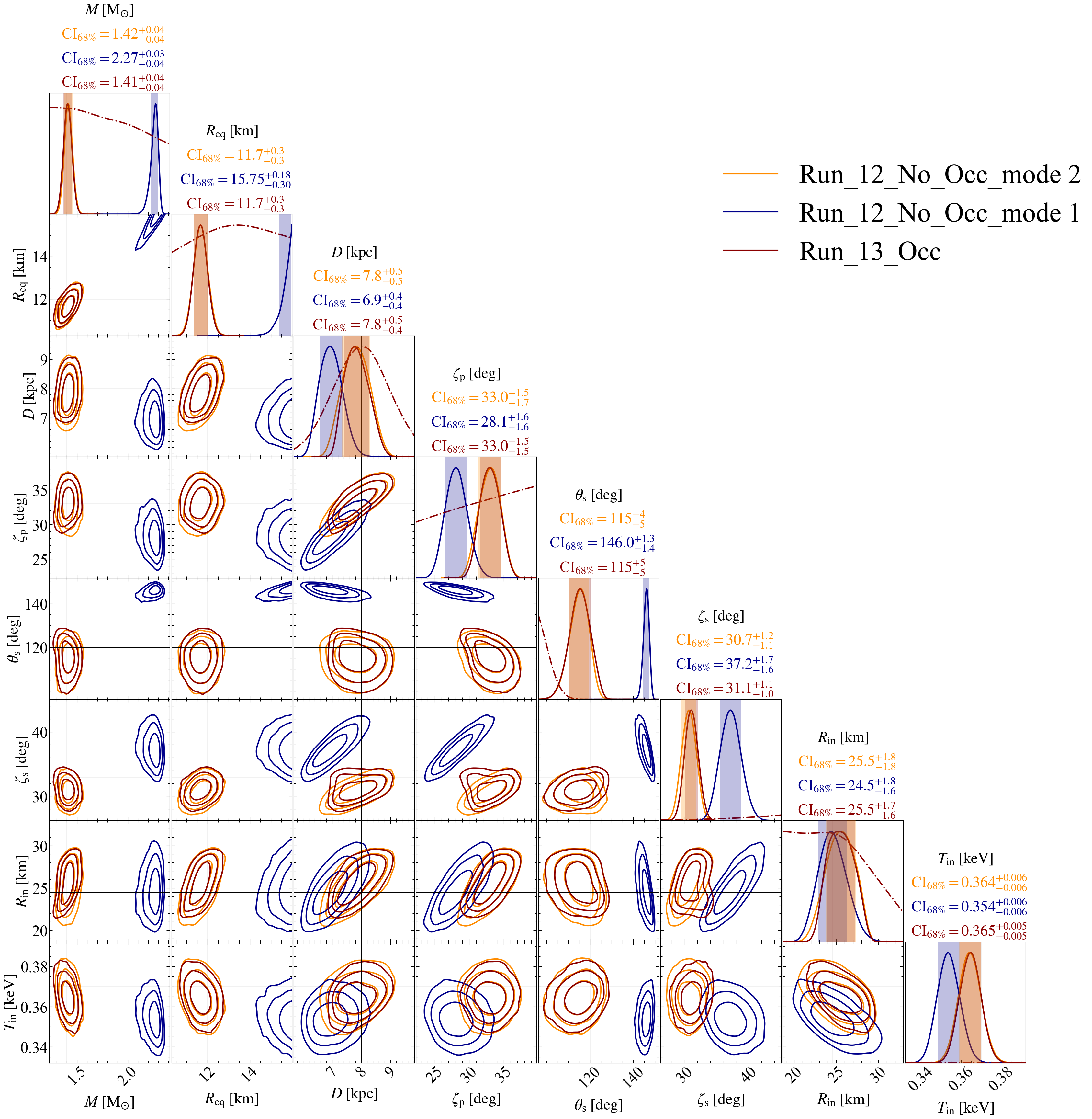}
    \caption{Posterior distributions for Runs 12 and 13 listed in Table~\ref{tab:sampling_runs_AB}. The yellow and blue colours correspond to the two modes obtained using the model without disc occultation, while the red colour represents the single solution obtained with the disc occultation model. The datasets are generated with disc occultation included in both cases.
}
    \label{fig:run_12_13_corner}
\end{figure*}

\begin{figure}
	\includegraphics[width=\columnwidth]{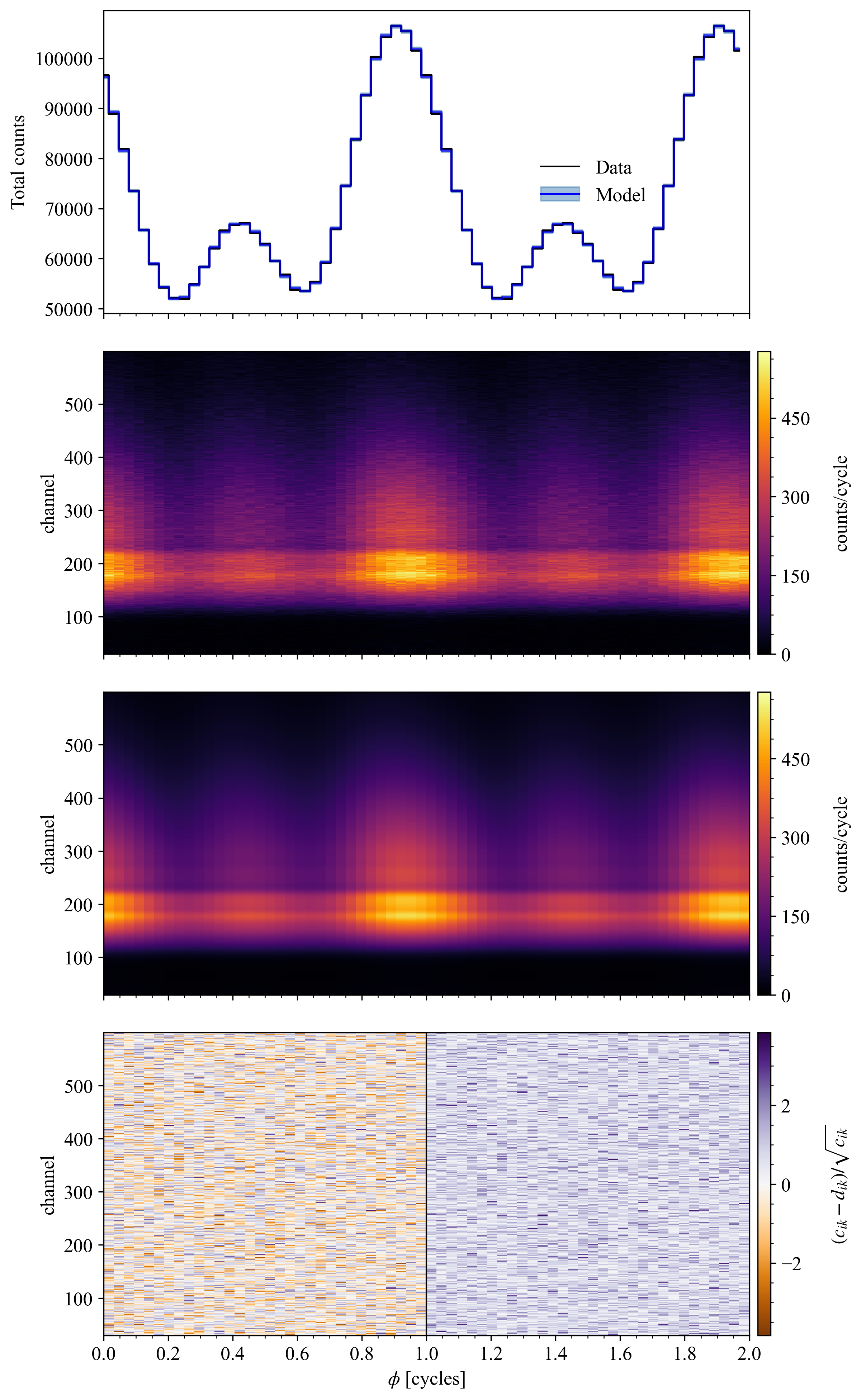}
    \caption{Same as Fig.~\ref{fig:run_3_4_residual}, but for Run 13.
}
    \label{fig:run_13_residual}
\end{figure}

\subsubsection{Inference results in Scenario B} \label{sec:inference of scenario B}
As reliable parameter recovery is hard to achieve in Scenario~A, we perform a series of additional runs for Scenario~B. These runs are summarised in Table~\ref{tab:sampling_runs_AB}, and the CI for each parameter is listed in the lower part of Table~\ref{tab:parameter CI}. Overall, Scenario~B represents a substantially more favourable configuration for parameter recovery, with most parameters recovered within the 68\% CI across all runs.

We begin with the simple configuration (Run~8), in which both $\theta_{\rm p}$ and $i$ are fixed and neither disc emission nor occultation is included. In this case, all 15 parameters are recovered within the 68\% CI, indicating highly accurate parameter recovery. When both disc emission and occultation are included (Runs~9 and 10), the recovery remains comparably robust, with only two parameters falling outside the 68\% CI.

For configurations with $\theta_{\rm p}$ free, the posterior shows more complex behaviour. In Run~11, disc occultation is included in the synthetic data but neglected in the sampling model. For this setup, we explore three different Poisson-noise realizations of the synthetic data, generated using random seeds 1, 2, and 42, and find that the inferred posteriors depend sensitively on the particular seed, with different choices leading to distinct solutions. To further investigate this behaviour, we repeat the same analysis using multimodal sampling in Run~12, which reveals two separated posterior modes (blue and yellow in Fig.~\ref{fig:run_12_13_corner}). The yellow mode closely matches the input parameters, and is consistent with the results obtained in Run~11 with seed 1. By contrast, the blue mode is strongly biased, yielding $M \simeq 2.2~{\rm M}_\odot$ and $R \simeq 15.7$~km, both significantly offset from the true values; this solution is consistent with the posteriors obtained in Run~11 with seeds 2 and 42.

For comparison, we also performed Run 13 using a model that includes disc occultation, again employing multimodal sampling. In this case, no secondary mode is found. The posterior distributions from this run are shown in red in Fig.~\ref{fig:run_12_13_corner}, overlap with mode 2 in Run~12. In these posteriors, all major parameters are well recovered, with both $M$ and $R_{\rm eq}$ constrained within 68\% CI, and with tight posterior constraints corresponding to $\sim 3\%$ precision in mass $\pm 0.04~{\rm M}_\odot$ and radius $\pm 0.3$ km.

Next, we consider Runs~14 and 15, in which all parameters are allowed to vary. For both the occultation and no-occultation models, we obtain good overall recovery of the input parameters. In particular, the no-occultation model does not exhibit a spurious secondary mode. A plausible interpretation is that allowing the inclination angle $i$ to vary reduces the geometric degeneracy present when $i$ is fixed, so that the compensating solution no longer appears as a distinct mode. Alternatively, when marginalised over $i$, the statistical weight of the spurious mode may become negligible, preventing it from being identified. From a practical perspective, fixing $i$ is unlikely to be realistic for real data, where it is typically constrained rather than fixed. A useful follow-up would therefore be to impose a tight prior on $i$, representative of realistic observational constraints, to test whether a similar spurious mode re-emerges. Such an investigation would help determine whether these modes are a genuine concern in real data analyses and remains to be explored in future work.

\section{Discussion} \label{sec:discussion}
In this section, we discuss the main implications of our results for parameter inference. Section~\ref{sec:Geometric Dependence of Parameter Recovery} examines how hotspot geometry affects the pulse profiles and parameter recovery, while Section~\ref{sec:Impact of Disc Occultation on Parameter Recovery} focuses on the role of disc occultation. Section~\ref{sec:Reliability of Parameter Recovery} then evaluates the overall reliability of the recovery. We also discuss the relevance of our findings for specific sources, including J1808 and J1444, in Section~\ref{sec:Implications for Disc Occultation in SAX J1808 and J1444}. Finally, Section~\ref{sec:Computational Cost of Including Disc Occultation} addresses the computational cost associated with including disc occultation.

\subsection{Geometric dependence of parameter recovery} \label{sec:Geometric Dependence of Parameter Recovery}

The difference in parameter recovery between Scenarios A and B appears to reflect differences in hotspot geometry and the resulting pulse profiles.
In Scenario A, the primary hotspot is located very close to the spin axis. From the observer’s perspective, this hotspot remains visible for nearly the entire rotational cycle, with only weak phase-dependent variations in its projected area. As a result, the pulse profile exhibits a single broad peak (see Fig.~\ref{fig:run_3_4_residual}). The contribution from the secondary hotspot overlaps with the flux minimum of the primary, reducing its depth and preventing the formation of a distinct peak. This polar-cap geometry therefore produces a single-peaked pulse profile with limited substructure.

In Scenario B, the hotspot is positioned at a moderate colatitude $\theta_{\rm p}=60\degr$ and $\theta_{\rm s}=120\degr$, with a moderate inclination $i=58\degr$. In this configuration, the hotspot moves into and out of the observer’s line of sight during a rotation, producing strong phase-dependent variations in its visible area. The secondary hotspot dominates at the opposite phase and forms a distinct peak. The resulting pulse profile therefore shows two well-separated peaks within a single rotation cycle (see the top panel of Fig.~\ref{fig:run_13_residual}).

These differences in pulse profile have important consequences for parameter recovery. The $\theta_{\rm p} = 60\degr$ configuration provides a richer set of observable features, including two peaks, their separation in phase, and their relative widths and amplitudes. These features may help to constrain multiple geometric parameters, including the hotspot colatitudes, sizes, and phase offsets, thereby reducing degeneracies. By contrast, the single-peaked pulse profile produced in Scenario A carries significantly less geometric information. Some combinations of parameters may generate similar pulse shapes. This may cause nested sampling to yield statistically acceptable fits while not stemming from the true input parameters. The difference in geometric identifiability is a plausible explanation for the more reliable parameter recovery observed in Scenario B compared to Scenario A.

\subsection{Impact of disc occultation on parameter recovery} \label{sec:Impact of Disc Occultation on Parameter Recovery}

Motivated by physical self-consistency, we introduce disc occultation into the model and perform a series of runs to assess whether its inclusion affects parameter recovery. In this subsection, we discuss the impact of disc occultation on parameter recovery, in particular its potential to bias the inference or alter the inferred solutions.

In Scenario A, we find that including or excluding occultation in the model leads to no clear qualitative change in the recovery results. 
In Scenario B, when the inclination angle is fixed, the model without occultation exhibits a spurious secondary mode, which disappears when all parameters are allowed to vary freely. Overall, the effect of disc occultation on parameter recovery appears to be present but relatively subtle. It remains unclear why the impact on parameter recovery is not stronger, despite the relatively large changes in the pulse profiles caused by occultation. Analyses of additional scenarios and parameter configurations may help to clarify the relationship between (non-)occulted pulse profiles and the success of parameter recovery.

Given that including occultation provides a more physically self-consistent description, we recommend incorporating it in future analyses of real data, provided that computational resources permit. As discussed in Section~\ref{sec:Computational Cost of Including Disc Occultation}, the associated increase in computational cost remains manageable, further supporting this recommendation.

\subsection{Reliability of parameter recovery} \label{sec:Reliability of Parameter Recovery}

In this subsection, we assess the robustness of parameter recovery based on the set of runs performed in this work. The \stu{} model is intrinsically complex, involving up to 19 free parameters when disc emission and occultation are included, and the robustness of parameter recovery in such a high-dimensional parameter space is therefore a central concern.

When the fits to the data are statistically acceptable and the parameter space is sufficiently well sampled, previous studies (e.g., \citealt{Bogdanov21,Holt2025}) have shown that pulse profile modelling can yield unbiased parameter constraints in the absence of an accretion disc. In our analysis, we also perform several runs without including a disc component, specifically Runs 1, 2, and 8. In all three cases, the inferred mass and radius are recovered within the 68\% CIs, when not separating the posteriors in different modes. These results appear consistent with previous findings.

The situation changes once the accretion disc is included. The addition of disc emission and occultation increases the overall model complexity and introduces additional free parameters. Under these conditions, the success of parameter recovery becomes more sensitive to the underlying hotspot geometry. For polar-cap configurations (Scenario A), the true values of $M$ and $R_{\rm eq}$ no longer lie within the 68\% CIs. By contrast, mid-colatitude geometry (Scenario B) allows for substantially more reliable recovery. This contrast suggests that in the presence of a disc, the robustness of parameter recovery may be influenced by the geometric configuration. However, this conclusion is based on only two representative scenarios, and a broader exploration of hotspot geometries will be necessary to assess its generality.

\begin{table*}
\centering
\small
\setlength{\tabcolsep}{6pt}
\caption{Typical runtimes (in hours) and sampling statistics for representative runs using 1000 live points and without multimodal sampling are presented. The \textit{Samples} column reports the total number of samples for each run, expressed in units of $10^6$. The \textit{Likelihood} column gives the average likelihood evaluation time. The \textit{Runtime} is reported in CPU core hours. All simulations were performed on the Dutch national supercomputer Snellius$^{\rm a}$. Each quantity is reported separately for models with and without disc occultation, enabling a direct comparison. The two cases use exactly the same synthetic datasets; the only difference lies in whether disc occultation is included in the sampling model.}
\label{tab:computational_times}

\begin{tabular}{@{}c c c c c c c c c@{}}
\hline 
Fix $i$ & Fix $\theta_{\rm p}$ & Scenario
& Samples (no occ) & Samples (occ)
& Likelihood (no occ) & Likelihood (occ)
& Runtime (no occ) & Runtime (occ) \\
& & 
& [$\times 10^6$] & [$\times 10^6$]
& [s] & [s]
& [hr] & [hr] \\
\hline

\checkmark & \checkmark & A
& $\sim 23$ & $\sim 17$
& $\sim 0.40$ & $\sim 0.43$
& $\sim 2500$ & $\sim 2100$ \\

\checkmark & $\times$ & A
& $\sim 31$ & $\sim 60$
& $\sim 0.48$ & $\sim 0.51$
& $\sim 4100$ & $\sim 8500$ \\

\checkmark & \checkmark & B
& $\sim 50$ & $\sim 77$
& $\sim 0.35$ & $\sim 0.41$
& $\sim 4800$ & $\sim 8700$ \\

\checkmark & $\times$ & B 
& $\sim 80$ & $\sim 99$
& $\sim 0.43$ & $\sim 0.36$
& $\sim 9400$ & $\sim 9700$ \\

 $\times$ &  $\times$ & B 
& $\sim 340$ & $\sim 650$
& $\sim 0.37$ & $\sim 0.35$
& $\sim 35000$ & $\sim 64000$ \\

\hline
\end{tabular}

\vspace{2pt}
\parbox{\textwidth}{\footnotesize
$^{\rm a}$ https://www.surf.nl/en/services/compute/snellius-the-national-supercomputer
}
\end{table*}

It should be noted that parameter inferences based on NICER data alone may not always be sufficiently robust to uniquely constrain complex geometric and emission models. This limitation could be alleviated through the inclusion of broader-band X-ray data or joint analyses combining NICER with other observatories. In particular, future missions such as \ac{eXTP} will provide enhanced spectral, timing, and polarimetric capabilities, offering valuable complementary constraints for PPM of AMPs \citep{Li25, Zhang2025}. Such multi-instrument approaches can help to reduce degeneracies and improve the reliability of parameter recovery in models that include an accretion disc.

\subsection{Implications for disc occultation in J1808 and J1444} \label{sec:Implications for Disc Occultation in SAX J1808 and J1444}

An important question arising from our analysis is whether the inferred geometries of J1808 and J1444 lie in a regime where disc occultation is expected to play a significant role.

For J1808, \citet{Dorsman2025} applied a model including disc emission to NICER observations of the 2019 and 2022 outbursts. Their posterior distributions place the primary hotspot close to the rotational pole, while the inclination spans a relatively broad range of approximately $30\degr$--$70\degr$. For J1444, in addition to the parameters adopted in Scenario A in this work, \citet{Papitto2025} carried out Stokes vector modelling, obtaining a best fit of $i = 74\fdg1$, $\theta_{\rm p} = 11\fdg8$, and $\theta_{\rm s} = 172\fdg6$. These parameter combinations imply a high-inclination configuration with a near-polar primary hotspot. In recent joint NICER and IXPE analyses of J1444, \citet{Dorsman2026} performed pulse profile modelling with disc occultation in \acs{X-PSI}, using the implementation introduced in this work, and obtained constraints on the system geometry consistent with this picture.

Based on these geometries, both sources are expected to lie in regimes where emission from the secondary hotspot can be partially obscured by the disc, as shown in Fig.~\ref{fig:Relative_Obscuration}. For J1808, the wide inclination range suggests that the system may lie near a transitional regime in which disc occultation may or may not be important, while for J1444, the higher inclination implies that disc blocking is likely to play a more significant role. For parameter recovery, as shown for Scenario A in Section~\ref{sec:Impact of Disc Occultation on Parameter Recovery}, including disc occultation has a limited impact on the posterior distributions. For J1808 and J1444, this suggests that neglecting occultation may not introduce significant biases, although its inclusion remains preferable for physical consistency.

Additional uncertainties may arise from the assumed standard thin disc model itself. For J1808, spectral analyses of NICER data using \acs{X-PSI} suggest that a simple multicolor disc model alone cannot fully reproduce the observed X-ray spectrum, and additional flexibility such as background marginalisation is often required \citep{Dorsman2025}. This indicates that the accretion flow or background emission is more complex than assumed in a standard thin disc picture, and motivates more realistic modelling of accretion emission as an important direction for future work.

\subsection{Computational cost of including disc occultation} \label{sec:Computational Cost of Including Disc Occultation}
The inclusion of disc occultation in the \stu{} model increases the computational cost of parameter inference, making it important to assess whether this additional computational overhead remains acceptable in practice. Table~\ref{tab:computational_times} summarises typical runtimes and sampling statistics for representative configurations in Scenarios A and B. The reported values correspond to runs performed without multimodal sampling and using 1000 live points.

Across the comparisons, the likelihood evaluation time in the occultation case does not show a consistent increase relative to the no-occultation case. The total runtime is primarily determined by the number of samples required to explore the parameter space. With disc occultation included, more samples are typically required. This is consistent with disc occultation introducing additional structure and complexity into the likelihood surface, which makes it more difficult to explore efficiently.

In the configurations explored here, disc occultation leads to lower compute time and fewer samples in the first comparison, while in the remaining comparisons it leads to increases in both the number of samples and the computational cost. In the last four cases, the inclusion of disc occultation increases the number of samples by approximately 20--90 per cent and the computational cost by approximately 3--110 per cent. Although this increase is sometimes not negligible, it remains moderate and does not constitute a prohibitive limitation for targeted parameter studies.

\section{Conclusions} \label{sec:conclusion}
In this work, we have extended \acs{X-PSI} by incorporating accretion disc occultation into the modelling of X-ray pulse profiles from AMPs. Within this framework, we combine a realistic NS atmosphere model, relativistic ray tracing, disc emission and occultation in a self-consistent pulse profile calculation.

We first explored the geometric dependence of disc occultation by quantifying relative obscuration across a broad parameter space. This analysis shows that disc occultation depends mainly on the inclination angle. For systems viewed at moderate to high inclinations, particularly above $i \simeq 50\degr$, disc blocking can suppress surface emission and significantly modify the pulse profiles. The hotspot colatitude also contributes, as obscuration begins once the hotspot extends into the southern hemisphere and increases progressively with colatitude. The inner disc radius and hotspot size have only a minor effect on the obscuration fraction. These results help define a criterion for when disc occultation should be included in pulse profile modelling and highlight the importance of independent constraints on inclination, for example from polarization measurements.

The impact of disc occultation on parameter recovery is examined through a series of sampling runs with two representative hotspot geometries. We demonstrate that the underlying hotspot configuration on its own already plays an important role in inference performance. Polar-cap geometry can be difficult to constrain, as statistically acceptable fits can still yield biased posteriors and poorly constrained hotspot parameters. Configurations with hotspots closer to the equatorial region produce pulse profiles with richer structure, which may enable more reliable recovery of the NS mass, radius, and hotspot properties. The impact of neglecting disc occultation was not uniform, ranging from negligible effects to the introduction of spurious posterior features. Meanwhile, the additional computation cost, due to increased complexity in the likelihood surface, appears to be within acceptable bounds. Overall, future analyses should include disc occultation and previous analyses that have not included it should be revisited.

This study is limited to two representative geometries, and the resulting conclusions are therefore not exhaustive. Future work should explore a broader range of hotspot geometries and develop more realistic emission and accretion flow models to better assess the reliability of NS parameter inference in realistic accretion scenarios.

\section*{Acknowledgements}
Y.-H. M. and X.-D. L. acknowledge support from the National Key Research and Development Program of China (2021YFA0718500) and the Natural Science Foundation of China under grant No. 12121003. A.L.W. acknowledges support from NWO grant ENW-XL OCENW.XL21.XL21.038 \textit{Probing the phase diagram of Quantum Chromodynamics} (PI: Watts). T.S. acknowledges funding by the Research Council of Finland grant 368807. T.S. and J.P. were partially supported by the Research Council of Finland Centre of Excellence in Neutron-Star Physics (grants 374063 and 374064).
We acknowledge NWO for providing access to Snellius, hosted by SURF through the Computing Time on National Computer Facilities call for proposals. We acknowledge extensive use of NASA’s Astrophysics Data System (ADS) Bibliographic Services and the ArXiv.

\section*{Data Availability}
The data underlying this work will be made publicly available on Zenodo (\href{https://zenodo.org/records/20642751}{10.5281/zenodo.20642751}) upon acceptance.



\bibliographystyle{mnras}
\bibliography{example} 

@ARTICLE{Nattila2017,
       author = {{N{\"a}ttil{\"a}}, J. and {Miller}, M.~C. and {Steiner}, A.~W. and {Kajava}, J.~J.~E. and {Suleimanov}, V.~F. and {Poutanen}, J.},
        title = "{Neutron star mass and radius measurements from atmospheric model fits to X-ray burst cooling tail spectra}",
      journal = {\aap},
         year = 2017,
        month = dec,
       volume = {608},
          eid = {A31},
        pages = {A31},
          doi = {10.1051/0004-6361/201731082},
archivePrefix = {arXiv},
       eprint = {1709.09120},
 primaryClass = {astro-ph.HE},
       adsurl = {https://ui.adsabs.harvard.edu/abs/2017A&A...608A..31N}
}

@ARTICLE{Molkov2024,
       author = {{Molkov}, Sergey V. and {Lutovinov}, Alexander A. and {Tsygankov}, Sergey S. and {Suleimanov}, Valery F. and {Poutanen}, Juri and {Lapshov}, Igor Yu. and {Mereminskiy}, Ilya A. and {Semena}, Andrei N. and {Arefiev}, Vadim A. and {Tkachenko}, Alexey Yu.},
        title = "{Discovery of SRGA J144459.2‑604207 with the SRG/ART-XC telescope: A well-tempered bursting accreting millisecond X-ray pulsar}",
      journal = {\aap},
         year = 2024,
        month = oct,
       volume = {690},
          eid = {A353},
        pages = {A353},
          doi = {10.1051/0004-6361/202450581},
archivePrefix = {arXiv},
       eprint = {2404.19709},
 primaryClass = {astro-ph.HE},
       adsurl = {https://ui.adsabs.harvard.edu/abs/2024A&A...690A.353M}
}

@ARTICLE{Ibragimov2009,
       author = {{Ibragimov}, Askar and {Poutanen}, Juri},
        title = "{Accreting millisecond pulsar SAX J1808.4-3658 during its 2002 outburst: evidence for a receding disc}",
      journal = {\mnras},
         year = 2009,
        month = nov,
       volume = {400},
       number = {1},
        pages = {492-508},
          doi = {10.1111/j.1365-2966.2009.15477.x},
archivePrefix = {arXiv},
       eprint = {0901.0073},
 primaryClass = {astro-ph.SR},
       adsurl = {https://ui.adsabs.harvard.edu/abs/2009MNRAS.400..492I}
}

@ARTICLE{Beloborodov2002,
       author = {{Beloborodov}, Andrei M.},
        title = "{Gravitational Bending of Light Near Compact Objects}",
      journal = {\apjl},
         year = 2002,
        month = feb,
       volume = {566},
       number = {2},
        pages = {L85-L88},
          doi = {10.1086/339511},
archivePrefix = {arXiv},
       eprint = {astro-ph/0201117},
 primaryClass = {astro-ph},
       adsurl = {https://ui.adsabs.harvard.edu/abs/2002ApJ...566L..85B}
}

@ARTICLE{Bobrikova2023,
       author = {{Bobrikova}, Anna and {Loktev}, Vladislav and {Salmi}, Tuomo and {Poutanen}, Juri},
        title = "{Polarized radiation from an accretion shock in accreting millisecond pulsars using exact Compton scattering formalism}",
      journal = {\aap},
         year = 2023,
        month = oct,
       volume = {678},
          eid = {A99},
        pages = {A99},
          doi = {10.1051/0004-6361/202346833},
archivePrefix = {arXiv},
       eprint = {2309.02329},
 primaryClass = {astro-ph.HE},
       adsurl = {https://ui.adsabs.harvard.edu/abs/2023A&A...678A..99B}
}

@ARTICLE{AlGendy2014,
       author = {{AlGendy}, Mohammad and {Morsink}, Sharon M.},
        title = "{Universality of the Acceleration due to Gravity on the Surface of a Rapidly Rotating Neutron Star}",
      journal = {\apj},
         year = 2014,
        month = aug,
       volume = {791},
       number = {2},
          eid = {78},
        pages = {78},
          doi = {10.1088/0004-637X/791/2/78},
archivePrefix = {arXiv},
       eprint = {1404.0609},
 primaryClass = {astro-ph.HE},
       adsurl = {https://ui.adsabs.harvard.edu/abs/2014ApJ...791...78A}
}

@ARTICLE{Morsink2007,
       author = {{Morsink}, Sharon M. and {Leahy}, Denis A. and {Cadeau}, Coire and {Braga}, John},
        title = "{The Oblate Schwarzschild Approximation for Light Curves of Rapidly Rotating Neutron Stars}",
      journal = {\apj},
         year = 2007,
        month = jul,
       volume = {663},
       number = {2},
        pages = {1244-1251},
          doi = {10.1086/518648},
archivePrefix = {arXiv},
       eprint = {astro-ph/0703123},
 primaryClass = {astro-ph},
       adsurl = {https://ui.adsabs.harvard.edu/abs/2007ApJ...663.1244M}
}

@ARTICLE{Miller1998,
       author = {{Miller}, M. Coleman and {Lamb}, Frederick K.},
        title = "{Bounds on the Compactness of Neutron Stars from Brightness Oscillations during X-Ray Bursts}",
      journal = {\apjl},
         year = 1998,
        month = may,
       volume = {499},
       number = {1},
        pages = {L37-L40},
          doi = {10.1086/311335},
archivePrefix = {arXiv},
       eprint = {astro-ph/9711325},
 primaryClass = {astro-ph},
       adsurl = {https://ui.adsabs.harvard.edu/abs/1998ApJ...499L..37M}
}

@ARTICLE{Poutanen2003,
       author = {{Poutanen}, Juri and {Gierli{\'n}ski}, Marek},
        title = "{On the nature of the X-ray emission from the accreting millisecond pulsar SAX J1808.4-3658}",
      journal = {\mnras},
         year = 2003,
        month = aug,
       volume = {343},
       number = {4},
        pages = {1301-1311},
          doi = {10.1046/j.1365-8711.2003.06773.x},
archivePrefix = {arXiv},
       eprint = {astro-ph/0303084},
 primaryClass = {astro-ph},
       adsurl = {https://ui.adsabs.harvard.edu/abs/2003MNRAS.343.1301P}
}

@ARTICLE{Mitsuda1984,
       author = {{Mitsuda}, K. and {Inoue}, H. and {Koyama}, K. and {Makishima}, K. and {Matsuoka}, M. and {Ogawara}, Y. and {Shibazaki}, N. and {Suzuki}, K. and {Tanaka}, Y. and {Hirano}, T.},
        title = "{Energy Spectra of Low-Mass Binary X-Ray Sources Observed from Tenma}",
      journal = {\pasj},
         year = 1984,
        month = dec,
       volume = {36},
       number = {4},
        pages = {741-759},
          doi = {10.1093/pasj/36.4.741},
       adsurl = {https://ui.adsabs.harvard.edu/abs/1984PASJ...36..741M}
}

@ARTICLE{Makishima1986,
       author = {{Makishima}, K. and {Maejima}, Y. and {Mitsuda}, K. and {Bradt}, H.~V. and {Remillard}, R.~A. and {Tuohy}, I.~R. and {Hoshi}, R. and {Nakagawa}, M.},
        title = "{Simultaneous X-Ray and Optical Observations of GX 339-4 in an X-Ray High State}",
      journal = {\apj},
         year = 1986,
        month = sep,
       volume = {308},
        pages = {635},
          doi = {10.1086/164534},
       adsurl = {https://ui.adsabs.harvard.edu/abs/1986ApJ...308..635M}
}

@ARTICLE{Wilms2000,
       author = {{Wilms}, J. and {Allen}, A. and {McCray}, R.},
        title = "{On the Absorption of X-Rays in the Interstellar Medium}",
      journal = {\apj},
         year = 2000,
        month = oct,
       volume = {542},
       number = {2},
        pages = {914-924},
          doi = {10.1086/317016},
archivePrefix = {arXiv},
       eprint = {astro-ph/0008425},
 primaryClass = {astro-ph},
       adsurl = {https://ui.adsabs.harvard.edu/abs/2000ApJ...542..914W}
}

@ARTICLE{Dorsman2025,
       author = {{Dorsman}, Bas and {Salmi}, Tuomo and {Watts}, Anna L. and {Ng}, Mason and {Kamath}, Satish and {Bobrikova}, Anna and {Poutanen}, Juri and {Loktev}, Vladislav and {Kini}, Yves and {Choudhury}, Devarshi and {Vinciguerra}, Serena and {Bogdanov}, Slavko and {Chakrabarty}, Deepto},
        title = "{Parameter constraints for accreting millisecond pulsars with synthetic NICER data}",
      journal = {\mnras},
         year = 2025,
        month = apr,
       volume = {538},
       number = {4},
        pages = {2853-2868},
          doi = {10.1093/mnras/staf438},
archivePrefix = {arXiv},
       eprint = {2409.07908},
 primaryClass = {astro-ph.HE},
       adsurl = {https://ui.adsabs.harvard.edu/abs/2025MNRAS.538.2853D}
}

@ARTICLE{Poutanen2006,
       author = {{Poutanen}, Juri and {Beloborodov}, Andrei M.},
        title = "{Pulse profiles of millisecond pulsars and their Fourier amplitudes}",
      journal = {\mnras},
         year = 2006,
        month = dec,
       volume = {373},
       number = {2},
        pages = {836-844},
          doi = {10.1111/j.1365-2966.2006.11088.x},
archivePrefix = {arXiv},
       eprint = {astro-ph/0608663},
 primaryClass = {astro-ph},
       adsurl = {https://ui.adsabs.harvard.edu/abs/2006MNRAS.373..836P}
}

@INPROCEEDINGS{Skilling2004,
       author = {{Skilling}, John},
        title = "{Nested Sampling}",
    booktitle = {Bayesian Inference and Maximum Entropy Methods in Science and Engineering: 24th International Workshop on Bayesian Inference and Maximum Entropy Methods in Science and Engineering},
         year = 2004,
       editor = {{Fischer}, Rainer and {Preuss}, Roland and {Toussaint}, Udo Von},
       series = {AIP Conf. Ser.},
       volume = {735},
        month = nov,
    publisher = {AIP},
address={Melville, NY},
        pages = {395-405},
          doi = {10.1063/1.1835238},
       adsurl = {https://ui.adsabs.harvard.edu/abs/2004AIPC..735..395S}
}

@ARTICLE{Feroz2009,
       author = {{Feroz}, F. and {Hobson}, M.~P. and {Bridges}, M.},
        title = "{MULTINEST: an efficient and robust Bayesian inference tool for cosmology and particle physics}",
      journal = {\mnras},
         year = 2009,
        month = oct,
       volume = {398},
       number = {4},
        pages = {1601-1614},
          doi = {10.1111/j.1365-2966.2009.14548.x},
archivePrefix = {arXiv},
       eprint = {0809.3437},
 primaryClass = {astro-ph},
       adsurl = {https://ui.adsabs.harvard.edu/abs/2009MNRAS.398.1601F}
}

@software{Buchner2016,
       author = {{Buchner}, Johannes},
        title = "{PyMultiNest: Python interface for MultiNest}",
 howpublished = {Astrophysics Source Code Library, record ascl:1606.005},
         year = 2016,
        month = jun,
          eid = {ascl:1606.005},
       adsurl = {https://ui.adsabs.harvard.edu/abs/2016ascl.soft06005B}
}

@ARTICLE{Feroz2019,
       author = {{Feroz}, Farhan and {Hobson}, Michael P. and {Cameron}, Ewan and {Pettitt}, Anthony N.},
        title = "{Importance Nested Sampling and the MultiNest Algorithm}",
      journal = {The Open Journal of Astrophysics},
         year = 2019,
        month = nov,
       volume = {2},
       number = {1},
          eid = {10},
        pages = {10},
          doi = {10.21105/astro.1306.2144},
archivePrefix = {arXiv},
       eprint = {1306.2144},
 primaryClass = {astro-ph.IM},
       adsurl = {https://ui.adsabs.harvard.edu/abs/2019OJAp....2E..10F}
}

@ARTICLE{Papitto2025,
       author = {{Papitto}, Alessandro and {Di Marco}, Alessandro and {Poutanen}, Juri and {Salmi}, Tuomo and {Illiano}, Giulia and {La Monaca}, Fabio and {Ambrosino}, Filippo and {Bobrikova}, Anna and {Baglio}, Maria Cristina and {Ballocco}, Caterina and {Burderi}, Luciano and {Campana}, Sergio and {Coti Zelati}, Francesco and {Di Salvo}, Tiziana and {La Placa}, Riccardo and {Loktev}, Vladislav and {Long}, Sinan and {Malacaria}, Christian and {Miraval Zanon}, Arianna and {Ng}, Mason and {Pilia}, Maura and {Sanna}, Andrea and {Stella}, Luigi and {Strohmayer}, Tod and {Zane}, Silvia},
        title = "{Discovery of polarized X-ray emission from the accreting millisecond pulsar SRGA J144459.2{\textendash}604207}",
      journal = {\aap},
         year = 2025,
        month = feb,
       volume = {694},
          eid = {A37},
        pages = {A37},
          doi = {10.1051/0004-6361/202451775},
archivePrefix = {arXiv},
       eprint = {2408.00608},
 primaryClass = {astro-ph.HE},
       adsurl = {https://ui.adsabs.harvard.edu/abs/2025A&A...694A..37P}
}

@ARTICLE{Ng2024,
       author = {{Ng}, Mason and {Ray}, Paul S. and {Sanna}, Andrea and {Strohmayer}, Tod E. and {Papitto}, Alessandro and {Illiano}, Giulia and {Albayati}, Arianna C. and {Altamirano}, Diego and {Boztepe}, Tu{\u{g}}ba and {G{\"u}ver}, Tolga and {Chakrabarty}, Deepto and {Arzoumanian}, Zaven and {Buisson}, D.~J.~K. and {Ferrara}, Elizabeth C. and {Gendreau}, Keith C. and {Guillot}, Sebastien and {Hare}, Jeremy and {Jaisawal}, Gaurava K. and {Malacaria}, Christian and {Wolff}, Michael T.},
        title = "{NICER Discovery that SRGA J144459.2{\textendash}604207 Is an Accreting Millisecond X-Ray Pulsar}",
      journal = {\apjl},
         year = 2024,
        month = jun,
       volume = {968},
       number = {1},
          eid = {L7},
        pages = {L7},
          doi = {10.3847/2041-8213/ad4edb},
archivePrefix = {arXiv},
       eprint = {2405.00087},
 primaryClass = {astro-ph.HE},
       adsurl = {https://ui.adsabs.harvard.edu/abs/2024ApJ...968L...7N}
}

@ARTICLE{Galloway2024,
       author = {{Galloway}, D.~K. and {Goodwin}, A.~J. and {Hilder}, T. and {Waterson}, L. and {Cup{\'a}k}, M.},
        title = "{Inferring system parameters from the bursts of the accretion-powered pulsar IGR J17498-2921}",
      journal = {\mnras},
         year = 2024,
        month = nov,
       volume = {535},
       number = {1},
        pages = {647-656},
          doi = {10.1093/mnras/stae2422},
archivePrefix = {arXiv},
       eprint = {2403.16471},
 primaryClass = {astro-ph.HE},
       adsurl = {https://ui.adsabs.harvard.edu/abs/2024MNRAS.535..647G}
}

@ARTICLE{Li25,
       author = {{Li}, Ang and {Watts}, Anna L. and {Zhang}, Guobao and {Guillot}, Sebastien and {Xu}, Yanjun and {Santangelo}, Andrea and {Zane}, Silvia and {Feng}, Hua and {Zhang}, Shuang-Nan and {Ge}, Mingyu and {Qi}, Liqiang and {Salmi}, Tuomo and {Dorsman}, Bas and {Miao}, Zhiqiang and {Tu}, Zhonghao and {Cavecchi}, Yuri and {Zhou}, Xia and {Zheng}, Xiaoping and {Wang}, Weihua and {Cheng}, Quan and {Liu}, Xuezhi and {Wei}, Yining and {Wang}, Wei and {Xu}, Yujing and {Weng}, Shanshan and {Zhu}, Weiwei and {Li}, Zhaosheng and {Shao}, Lijing and {Tuo}, Youli and {Dohi}, Akira and {Lyu}, Ming and {Liu}, Peng and {Yuan}, Jianping and {Wang}, Mingyang and {Zhang}, Wenda and {Li}, Zexi and {Tao}, Lian and {Zhang}, Liang and {Shen}, Hong and {Provid{\^e}ncia}, Constan{\c{c}}a and {Tolos}, Laura and {Patruno}, Alessandro and {Li}, Li and {Liu}, Guozhu and {Zhou}, Kai and {Chen}, Lie-Wen and {Fan}, Yizhong and {Kajino}, Toshitaka and {Lai}, Dong and {Li}, Xiangdong and {Meng}, Jie and {Tang}, Xiaodong and {Xiao}, Zhigang and {Xiong}, Shaolin and {Xu}, Renxin and {Zhou}, Shan-Gui and {Ballantyne}, David R. and {Burgio}, G. Fiorella and {Chenevez}, J{\'e}r{\^o}me and {Choudhury}, Devarshi and {Fantina}, Anthea F. and {Galloway}, Duncan K. and {Gulminelli}, Francesca and {Hebeler}, Kai and {Hoogkamer}, Mariska and {Horvath}, Jorge E. and {Kini}, Yves and {Kurkela}, Aleksi and {Linares}, Manuel and {Margueron}, J{\'e}r{\^o}me and {Mendes}, Melissa and {Oertel}, Micaela and {Papitto}, Alessandro and {Poutanen}, Juri and {Rea}, Nanda and {Schwenk}, Achim and {Song}, Xin-Ying and {Svensson}, Isak and {Tsang}, David and {Vuorinen}, Aleksi and {Andersson}, Nils and {Miller}, M. Coleman and {Rezzolla}, Luciano and {Stone}, Jirina R. and {Thomas}, Anthony W.},
        title = "{Dense matter in neutron stars with eXTP}",
      journal = {Science China Physics, Mechanics, and Astronomy},
         year = 2025,
        month = sep,
       volume = {68},
       number = {11},
          eid = {119503},
        pages = {119503},
          doi = {10.1007/s11433-025-2761-4},
archivePrefix = {arXiv},
       eprint = {2506.08104},
 primaryClass = {astro-ph.HE},
       adsurl = {https://ui.adsabs.harvard.edu/abs/2025SCPMA..6819503L}
}

@INPROCEEDINGS{Gendreau16,
       author = {{Gendreau}, Keith C. and {Arzoumanian}, Zaven and {Adkins}, Phillip W. and {Albert}, Cheryl L. and {Anders}, John F. and {Aylward}, Andrew T. and {Baker}, Charles L. and {Balsamo}, Erin R. and {Bamford}, William A. and {Benegalrao}, Suyog S. and {Berry}, Daniel L. and {Bhalwani}, Shiraz and {Black}, J. Kevin and {Blaurock}, Carl and {Bronke}, Ginger M. and {Brown}, Gary L. and {Budinoff}, Jason G. and {Cantwell}, Jeffrey D. and {Cazeau}, Thoniel and {Chen}, Philip T. and {Clement}, Thomas G. and {Colangelo}, Andrew T. and {Coleman}, Jerry S. and {Coopersmith}, Jonathan D. and {Dehaven}, William E. and {Doty}, John P. and {Egan}, Mark D. and {Enoto}, Teruaki and {Fan}, Terry W. and {Ferro}, Deneen M. and {Foster}, Richard and {Galassi}, Nicholas M. and {Gallo}, Luis D. and {Green}, Chris M. and {Grosh}, Dave and {Ha}, Kong Q. and {Hasouneh}, Monther A. and {Heefner}, Kristofer B. and {Hestnes}, Phyllis and {Hoge}, Lisa J. and {Jacobs}, Tawanda M. and {J{\o}rgensen}, John L. and {Kaiser}, Michael A. and {Kellogg}, James W. and {Kenyon}, Steven J. and {Koenecke}, Richard G. and {Kozon}, Robert P. and {LaMarr}, Beverly and {Lambertson}, Mike D. and {Larson}, Anne M. and {Lentine}, Steven and {Lewis}, Jesse H. and {Lilly}, Michael G. and {Liu}, Kuochia Alice and {Malonis}, Andrew and {Manthripragada}, Sridhar S. and {Markwardt}, Craig B. and {Matonak}, Bryan D. and {Mcginnis}, Isaac E. and {Miller}, Roger L. and {Mitchell}, Alissa L. and {Mitchell}, Jason W. and {Mohammed}, Jelila S. and {Monroe}, Charles A. and {Montt de Garcia}, Kristina M. and {Mul{\'e}}, Peter D. and {Nagao}, Louis T. and {Ngo}, Son N. and {Norris}, Eric D. and {Norwood}, Dwight A. and {Novotka}, Joseph and {Okajima}, Takashi and {Olsen}, Lawrence G. and {Onyeachu}, Chimaobi O. and {Orosco}, Henry Y. and {Peterson}, Jacqualine R. and {Pevear}, Kristina N. and {Pham}, Karen K. and {Pollard}, Sue E. and {Pope}, John S. and {Powers}, Daniel F. and {Powers}, Charles E. and {Price}, Samuel R. and {Prigozhin}, Gregory Y. and {Ramirez}, Julian B. and {Reid}, Winston J. and {Remillard}, Ronald A. and {Rogstad}, Eric M. and {Rosecrans}, Glenn P. and {Rowe}, John N. and {Sager}, Jennifer A. and {Sanders}, Claude A. and {Savadkin}, Bruce and {Saylor}, Maxine R. and {Schaeffer}, Alexander F. and {Schweiss}, Nancy S. and {Semper}, Sean R. and {Serlemitsos}, Peter J. and {Shackelford}, Larry V. and {Soong}, Yang and {Struebel}, Jonathan and {Vezie}, Michael L. and {Villasenor}, Joel S. and {Winternitz}, Luke B. and {Wofford}, George I. and {Wright}, Michael R. and {Yang}, Mike Y. and {Yu}, Wayne H.},
        title = "{The Neutron star Interior Composition Explorer (NICER): design and development}",
    booktitle = {Space Telescopes and Instrumentation 2016: Ultraviolet to Gamma Ray},
         year = 2016,
       editor = {{den Herder}, Jan-Willem A. and {Takahashi}, Tadayuki and {Bautz}, Marshall},
       series = {\procspie},
       volume = {9905},
        month = jul,
          eid = {99051H},
        pages = {99051H},
          doi = {10.1117/12.2231304},
       adsurl = {https://ui.adsabs.harvard.edu/abs/2016SPIE.9905E..1HG}
}

@ARTICLE{Pechenick83,
       author = {{Pechenick}, K.~R. and {Ftaclas}, C. and {Cohen}, J.~M.},
        title = "{Hot spots on neutron stars - The near-field gravitational lens}",
      journal = {\apj},
         year = 1983,
        month = nov,
       volume = {274},
        pages = {846-857},
          doi = {10.1086/161498},
       adsurl = {https://ui.adsabs.harvard.edu/abs/1983ApJ...274..846P}
}

@ARTICLE{Bogdanov19b,
       author = {{Bogdanov}, Slavko and {Lamb}, Frederick K. and {Mahmoodifar}, Simin and {Miller}, M. Coleman and {Morsink}, Sharon M. and {Riley}, Thomas E. and {Strohmayer}, Tod E. and {Tung}, Albert K. and {Watts}, Anna L. and {Dittmann}, Alexander J. and {Chakrabarty}, Deepto and {Guillot}, Sebastien and {Arzoumanian}, Zaven and {Gendreau}, Keith C.},
        title = "{Constraining the Neutron Star Mass-Radius Relation and Dense Matter Equation of State with NICER. II. Emission from Hot Spots on a Rapidly Rotating Neutron Star}",
      journal = {\apjl},
         year = 2019,
        month = dec,
       volume = {887},
       number = {1},
          eid = {L26},
        pages = {L26},
          doi = {10.3847/2041-8213/ab5968},
archivePrefix = {arXiv},
       eprint = {1912.05707},
 primaryClass = {astro-ph.HE},
       adsurl = {https://ui.adsabs.harvard.edu/abs/2019ApJ...887L..26B}
}

@ARTICLE{Bogdanov21,
       author = {{Bogdanov}, Slavko and {Dittmann}, Alexander J. and {Ho}, Wynn C.~G. and {Lamb}, Frederick K. and {Mahmoodifar}, Simin and {Miller}, M. Coleman and {Morsink}, Sharon M. and {Riley}, Thomas E. and {Strohmayer}, Tod E. and {Watts}, Anna L. and {Choudhury}, Devarshi and {Guillot}, Sebastien and {Harding}, Alice K. and {Ray}, Paul S. and {Wadiasingh}, Zorawar and {Wolff}, Michael T. and {Markwardt}, Craig B. and {Arzoumanian}, Zaven and {Gendreau}, Keith C.},
        title = "{Constraining the Neutron Star Mass-Radius Relation and Dense Matter Equation of State with NICER. III. Model Description and Verification of Parameter Estimation Codes}",
      journal = {\apjl},
         year = 2021,
        month = jun,
       volume = {914},
       number = {1},
          eid = {L15},
        pages = {L15},
          doi = {10.3847/2041-8213/abfb79},
archivePrefix = {arXiv},
       eprint = {2104.06928},
 primaryClass = {astro-ph.HE},
       adsurl = {https://ui.adsabs.harvard.edu/abs/2021ApJ...914L..15B}
}

@ARTICLE{Chatziioannou25,
       author = {{Chatziioannou}, Katerina and {Cromartie}, H. Thankful and {Gandolfi}, Stefano and {Tews}, Ingo and {Radice}, David and {Steiner}, Andrew W. and {Watts}, Anna L.},
        title = "{Neutron stars and the dense matter equation of state}",
      journal = {Reviews of Modern Physics},
         year = 2025,
        month = oct,
       volume = {97},
       number = {4},
          eid = {045007},
        pages = {045007},
          doi = {10.1103/ymsq-cfcw},
archivePrefix = {arXiv},
       eprint = {2407.11153},
 primaryClass = {nucl-th},
       adsurl = {https://ui.adsabs.harvard.edu/abs/2025RvMP...97d5007C}
}

@ARTICLE{Choudhury24,
       author = {{Choudhury}, Devarshi and {Salmi}, Tuomo and {Vinciguerra}, Serena and {Riley}, Thomas E. and {Kini}, Yves and {Watts}, Anna L. and {Dorsman}, Bas and {Bogdanov}, Slavko and {Guillot}, Sebastien and {Ray}, Paul S. and {Reardon}, Daniel J. and {Remillard}, Ronald A. and {Bilous}, Anna V. and {Huppenkothen}, Daniela and {Lattimer}, James M. and {Rutherford}, Nathan and {Arzoumanian}, Zaven and {Gendreau}, Keith C. and {Morsink}, Sharon M. and {Ho}, Wynn C.~G.},
        title = "{A NICER View of the Nearest and Brightest Millisecond Pulsar: PSR J0437─4715}",
      journal = {\apjl},
         year = 2024,
        month = aug,
       volume = {971},
       number = {1},
          eid = {L20},
        pages = {L20},
          doi = {10.3847/2041-8213/ad5a6f},
archivePrefix = {arXiv},
       eprint = {2407.06789},
 primaryClass = {astro-ph.HE},
       adsurl = {https://ui.adsabs.harvard.edu/abs/2024ApJ...971L..20C}
}

@ARTICLE{Salmi24,
       author = {{Salmi}, Tuomo and {Choudhury}, Devarshi and {Kini}, Yves and {Riley}, Thomas E. and {Vinciguerra}, Serena and {Watts}, Anna L. and {Wolff}, Michael T. and {Arzoumanian}, Zaven and {Bogdanov}, Slavko and {Chakrabarty}, Deepto and {Gendreau}, Keith and {Guillot}, Sebastien and {Ho}, Wynn C.~G. and {Huppenkothen}, Daniela and {Ludlam}, Renee M. and {Morsink}, Sharon M. and {Ray}, Paul S.},
        title = "{The Radius of the High-mass Pulsar PSR J0740+6620 with 3.6 yr of NICER Data}",
      journal = {\apj},
         year = 2024,
        month = oct,
       volume = {974},
       number = {2},
          eid = {294},
        pages = {294},
          doi = {10.3847/1538-4357/ad5f1f},
archivePrefix = {arXiv},
       eprint = {2406.14466},
 primaryClass = {astro-ph.HE},
       adsurl = {https://ui.adsabs.harvard.edu/abs/2024ApJ...974..294S}
}

@ARTICLE{Mauviard25,
       author = {{Mauviard}, Lucien and {Guillot}, Sebastien and {Salmi}, Tuomo and {Choudhury}, Devarshi and {Dorsman}, Bas and {Gonz{\'a}lez-Caniulef}, Denis and {Hoogkamer}, Mariska and {Huppenkothen}, Daniela and {Kazantsev}, Christine and {Kini}, Yves and {Olive}, Jean-Francois and {Stammler}, Pierre and {Watts}, Anna L. and {Mendes}, Melissa and {Rutherford}, Nathan and {Schwenk}, Achim and {Svensson}, Isak and {Bogdanov}, Slavko and {Kerr}, Matthew and {Ray}, Paul S. and {Guillemot}, Lucas and {Cognard}, Isma{\"e}l and {Theureau}, Gilles},
        title = "{A NICER View of the 1.4 M$_{{\ensuremath{\odot}}}$ Edge-on Pulsar PSR J0614-3329}",
      journal = {\apj},
         year = 2025,
        month = dec,
       volume = {995},
       number = {1},
          eid = {60},
        pages = {60},
          doi = {10.3847/1538-4357/ae145d},
archivePrefix = {arXiv},
       eprint = {2506.14883},
 primaryClass = {astro-ph.HE},
       adsurl = {https://ui.adsabs.harvard.edu/abs/2025ApJ...995...60M}
}

@ARTICLE{Dittmann24,
       author = {{Dittmann}, Alexander J. and {Miller}, M. Coleman and {Lamb}, Frederick K. and {Holt}, Isiah M. and {Chirenti}, Cecilia and {Wolff}, Michael T. and {Bogdanov}, Slavko and {Guillot}, Sebastien and {Ho}, Wynn C.~G. and {Morsink}, Sharon M. and {Arzoumanian}, Zaven and {Gendreau}, Keith C.},
        title = "{A More Precise Measurement of the Radius of PSR J0740+6620 Using Updated NICER Data}",
      journal = {\apj},
         year = 2024,
        month = oct,
       volume = {974},
       number = {2},
          eid = {295},
        pages = {295},
          doi = {10.3847/1538-4357/ad5f1e},
archivePrefix = {arXiv},
       eprint = {2406.14467},
 primaryClass = {astro-ph.HE},
       adsurl = {https://ui.adsabs.harvard.edu/abs/2024ApJ...974..295D}
}

@ARTICLE{Salmi18,
       author = {{Salmi}, T. and {N{\"a}ttil{\"a}}, J. and {Poutanen}, J.},
        title = "{Bayesian parameter constraints for neutron star masses and radii using X-ray timing observations of accretion-powered millisecond pulsars}",
      journal = {\aap},
         year = 2018,
        month = oct,
       volume = {618},
          eid = {A161},
        pages = {A161},
          doi = {10.1051/0004-6361/201833348},
archivePrefix = {arXiv},
       eprint = {1805.01149},
 primaryClass = {astro-ph.HE},
       adsurl = {https://ui.adsabs.harvard.edu/abs/2018A&A...618A.161S}
}

@INPROCEEDINGS{Patruno21,
       author = {{Patruno}, Alessandro and {Watts}, Anna L.},
        title = "{Accreting Millisecond X-ray Pulsars}",
    booktitle = {Timing Neutron Stars: Pulsations, Oscillations and Explosions},
         year = 2021,
       editor = {{Belloni}, Tomaso M. and {M{\'e}ndez}, Mariano and {Zhang}, Chengmin},
       series = {ASSL},
       volume = {461},
        month = jan,
        pages = {143-208},
          doi = {10.1007/978-3-662-62110-3_4},
archivePrefix = {arXiv},
       eprint = {1206.2727},
 primaryClass = {astro-ph.HE},
       adsurl = {https://ui.adsabs.harvard.edu/abs/2021ASSL..461..143P}
}

@INPROCEEDINGS{DiSalvo22,
       author = {{Di Salvo}, Tiziana and {Sanna}, Andrea},
        title = "{Accretion Powered X-ray Millisecond Pulsars}",
    booktitle = {Millisecond Pulsars},
         year = 2022,
       editor = {{Bhattacharyya}, Sudip and {Papitto}, Alessandro and {Bhattacharya}, Dipankar},
       series = {ASSL},
       volume = {465},
        month = jan,
        pages = {87-124},
          doi = {10.1007/978-3-030-85198-9_4},
       adsurl = {https://ui.adsabs.harvard.edu/abs/2022ASSL..465...87D}
}

@ARTICLE{Viironen04,
       author = {{Viironen}, K. and {Poutanen}, J.},
        title = "{Light curves and polarization of accretion- and nuclear-powered millisecond pulsars}",
      journal = {\aap},
         year = 2004,
        month = nov,
       volume = {426},
        pages = {985-997},
          doi = {10.1051/0004-6361:20041084},
archivePrefix = {arXiv},
       eprint = {astro-ph/0408250},
 primaryClass = {astro-ph},
       adsurl = {https://ui.adsabs.harvard.edu/abs/2004A&A...426..985V}
}

@article{IXPE2022,
       author = {{Weisskopf}, Martin C. and {Soffitta}, Paolo and {Baldini}, Luca and {Ramsey}, Brian D. and {O'Dell}, Stephen L. and {Romani}, Roger W. and {Matt}, Giorgio and {Deininger}, William D. and {Baumgartner}, Wayne H. and {Bellazzini}, Ronaldo and {Costa}, Enrico and {Kolodziejczak}, Jeffery J. and {Latronico}, Luca and {Marshall}, Herman L. and {Muleri}, Fabio and {Bongiorno}, Stephen D. and {Tennant}, Allyn and {Bucciantini}, Niccolo and {Dovciak}, Michal and {Marin}, Frederic and {Marscher}, Alan and {Poutanen}, Juri and {Slane}, Pat and {Turolla}, Roberto and {Kalinowski}, William and {Di Marco}, Alessandro and {Fabiani}, Sergio and {Minuti}, Massimo and {La Monaca}, Fabio and {Pinchera}, Michele and {Rankin}, John and {Sgro'}, Carmelo and {Trois}, Alessio and {Xie}, Fei and {Alexander}, Cheryl and {Allen}, D. Zachery and {Amici}, Fabrizio and {Andersen}, Jason and {Antonelli}, Angelo and {Antoniak}, Spencer and {Attina'}, Primo and {Barbanera}, Mattia and {Bachetti}, Matteo and {Baggett}, Randy M. and {Bladt}, Jeff and {Brez}, Alessandro and {Bonino}, Raffaella and {Boree}, Christopher and {Borotto}, Fabio and {Breeding}, Shawn and {Brienza}, Daniele and {Bygott}, H. Kyle and {Caporale}, Ciro and {Cardelli}, Claudia and {Carpentiero}, Rita and {Castellano}, Simone and {Castronuovo}, Marco and {Cavalli}, Luca and {Cavazzuti}, Elisabetta and {Ceccanti}, Marco and {Centrone}, Mauro and {Citraro}, Saverio and {D' Amico}, Fabio and {D'Alba}, Elisa and {Di Gesu}, Laura and {Del Monte}, Ettore and {Dietz}, Kurtis L. and {Di Lalla}, Niccolo' and {Di Persio}, Giuseppe and {Dolan}, David and {Donnarumma}, Immacolata and {Evangelista}, Yuri and {Ferrant}, Kevin and {Ferrazzoli}, Riccardo and {Ferrie}, MacKenzie and {Footdale}, Joseph and {Forsyth}, Brent and {Foster}, Michelle and {Garelick}, Benjamin and {Gunji}, Shuichi and {Gurnee}, Eli and {Head}, Michael and {Hibbard}, Grant and {Johnson}, Samantha and {Kelly}, Erik and {Kilaru}, Kiranmayee and {Lefevre}, Carlo and {Le Roy}, Shelley and {Loffredo}, Pasqualino and {Lorenzi}, Paolo and {Lucchesi}, Leonardo and {Maddox}, Tyler and {Magazzu}, Guido and {Maldera}, Simone and {Manfreda}, Alberto and {Mangraviti}, Elio and {Marengo}, Marco and {Marrocchesi}, Alessandra and {Massaro}, Francesco and {Mauger}, David and {McCracken}, Jeffrey and {McEachen}, Michael and {Mize}, Rondal and {Mereu}, Paolo and {Mitchell}, Scott and {Mitsuishi}, Ikuyuki and {Morbidini}, Alfredo and {Mosti}, Federico and {Nasimi}, Hikmat and {Negri}, Barbara and {Negro}, Michela and {Nguyen}, Toan and {Nitschke}, Isaac and {Nuti}, Alessio and {Onizuka}, Mitch and {Oppedisano}, Chiara and {Orsini}, Leonardo and {Osborne}, Darren and {Pacheco}, Richard and {Paggi}, Alessandro and {Painter}, Will and {Pavelitz}, Steven D. and {Pentz}, Christina and {Piazzolla}, Raffaele and {Perri}, Matteo and {Pesce-Rollins}, Melissa and {Peterson}, Colin and {Pilia}, Maura and {Profeti}, Alessandro and {Puccetti}, Simonetta and {Ranganathan}, Jaganathan and {Ratheesh}, Ajay and {Reedy}, Lee and {Root}, Noah and {Rubini}, Alda and {Ruswick}, Stephanie and {Sanchez}, Javier and {Sarra}, Paolo and {Santoli}, Francesco and {Scalise}, Emanuele and {Sciortino}, Andrea and {Schroeder}, Christopher and {Seek}, Tim and {Sosdian}, Kalie and {Spandre}, Gloria and {Speegle}, Chet O. and {Tamagawa}, Toru and {Tardiola}, Marcello and {Tobia}, Antonino and {Thomas}, Nicholas E. and {Valerie}, Robert and {Vimercati}, Marco and {Walden}, Amy L. and {Weddendorf}, Bruce and {Wedmore}, Jeffrey and {Welch}, David and {Zanetti}, Davide and {Zanetti}, Francesco},
        title = "{The Imaging X-Ray Polarimetry Explorer (IXPE): Pre-Launch}",
      journal = {JATIS},
         year = 2022,
        month = apr,
       volume = {8},
       number = {2},
       pages = {026002},
          doi = {10.1117/1.JATIS.8.2.026002},
archivePrefix = {arXiv},
       eprint = {2112.01269},
 primaryClass = {astro-ph.IM},
       adsurl = {https://ui.adsabs.harvard.edu/abs/2021arXiv211201269W}
}

@ARTICLE{Poutanen20,
       author = {{Poutanen}, Juri},
        title = "{Relativistic rotating vector model for X-ray millisecond pulsars}",
      journal = {\aap},
         year = 2020,
        month = sep,
       volume = {641},
          eid = {A166},
        pages = {A166},
          doi = {10.1051/0004-6361/202038689},
archivePrefix = {arXiv},
       eprint = {2006.10448},
 primaryClass = {astro-ph.HE},
       adsurl = {https://ui.adsabs.harvard.edu/abs/2020A&A...641A.166P}
}

@ARTICLE{Riley23,
       author = {{Riley}, Thomas E. and {Choudhury}, Devarshi and {Salmi}, Tuomo and {Vinciguerra}, Serena and {Kini}, Yves and {Dorsman}, Bas and {Watts}, Anna L. and {Huppenkothen}, Daniela and {Guillot}, Sebastien},
        title = "{X-PSI: A Python package for neutron star X-ray pulse simulation and inference}",
      journal = {The Journal of Open Source Software},
         year = 2023,
        month = feb,
       volume = {8},
       number = {82},
          eid = {4977},
        pages = {4977},
          doi = {10.21105/joss.04977},
       adsurl = {https://ui.adsabs.harvard.edu/abs/2023JOSS....8.4977R}
}

@ARTICLE{Salmi21,
       author = {{Salmi}, Tuomo and {Loktev}, Vladislav and {Korsman}, Karri and {Baldini}, Luca and {Tsygankov}, Sergey S. and {Poutanen}, Juri},
        title = "{Neutron star parameter constraints for accretion-powered millisecond pulsars from the simulated IXPE data}",
      journal = {\aap},
         year = 2021,
        month = feb,
       volume = {646},
          eid = {A23},
        pages = {A23},
          doi = {10.1051/0004-6361/202039470},
archivePrefix = {arXiv},
       eprint = {2009.09744},
 primaryClass = {astro-ph.HE},
       adsurl = {https://ui.adsabs.harvard.edu/abs/2021A&A...646A..23S}
}

@ARTICLE{Salmi25,
       author = {{Salmi}, Tuomo and {Dorsman}, Bas and {Watts}, Anna L. and {Bobrikova}, Anna and {Marco}, Alessandro Di and {Loktev}, Vladislav and {Papitto}, Alessandro and {Pilia}, Maura and {Poutanen}, Juri and {Rankin}, John},
        title = "{Modelling polarized X-ray pulses from accreting millisecond pulsars with X-PSI, using different hot spot locations and shapes}",
      journal = {\mnras},
         year = 2025,
        month = apr,
       volume = {538},
       number = {4},
        pages = {2562-2568},
          doi = {10.1093/mnras/staf441},
archivePrefix = {arXiv},
       eprint = {2501.12190},
 primaryClass = {astro-ph.HE},
       adsurl = {https://ui.adsabs.harvard.edu/abs/2025MNRAS.538.2562S}
}

@ARTICLE{Dorsman26,
       author = {{Dorsman}, Bas and {Salmi}, Tuomo and {Watts}, Anna L. and {Ng}, Mason and {Bobrikova}, Anna and {Loktev}, Vladislav and {Poutanen}, Juri and {Wilms}, Joern},
        title = "{Pulse profile modelling of the accretion-powered millisecond pulsar SAX J1808.4{\ensuremath{-}}3658 using NICER data from its 2019 and 2022 outbursts}",
      journal = {\mnras},
         year = 2026,
        month = jan,
       volume = {545},
       number = {2},
          eid = {staf1983},
        pages = {staf1983},
          doi = {10.1093/mnras/staf1983},
archivePrefix = {arXiv},
       eprint = {2511.07152},
 primaryClass = {astro-ph.HE},
       adsurl = {https://ui.adsabs.harvard.edu/abs/2026MNRAS.545f1983D}
}

@ARTICLE{Malacaria2025,
       author = {{Malacaria}, Christian and {Papitto}, Alessandro and {Campana}, Sergio and {Di Marco}, Alessandro and {Di Salvo}, Tiziana and {Cristina Baglio}, Maria and {Illiano}, Giulia and {La Placa}, Riccardo and {Miraval Zanon}, Arianna and {Pilia}, Maura and {Poutanen}, Juri and {Salmi}, Tuomo and {Sanna}, Andrea and {Mandal}, Manoj},
        title = "{Disk reflection and energetics from the accreting millisecond pulsar SRGA J144459.2‑604207}",
      journal = {\aap},
         year = 2025,
        month = jul,
       volume = {699},
          eid = {A288},
        pages = {A288},
          doi = {10.1051/0004-6361/202554075},
archivePrefix = {arXiv},
       eprint = {2502.08239},
 primaryClass = {astro-ph.HE},
       adsurl = {https://ui.adsabs.harvard.edu/abs/2025A&A...699A.288M}
}

@ARTICLE{Li2023,
       author = {{Li}, Zhaosheng and {Kuiper}, Lucien and {Ge}, Mingyu and {Falanga}, Maurizio and {Poutanen}, Juri and {Ji}, Long and {Pan}, Yuanyue and {Huang}, Yue and {Xu}, Renxin and {Song}, Liming and {Qu}, Jinlu and {Zhang}, Shu and {Lu}, Fangjun and {Zhang}, Shuang-Nan},
        title = "{Broadband X-Ray Timing and Spectral Characteristics of the Accretion-powered Millisecond X-Ray Pulsar MAXIJ1816-195}",
      journal = {\apj},
         year = 2023,
        month = dec,
       volume = {958},
       number = {2},
          eid = {177},
        pages = {177},
          doi = {10.3847/1538-4357/ad0296},
archivePrefix = {arXiv},
       eprint = {2303.11603},
 primaryClass = {astro-ph.HE},
       adsurl = {https://ui.adsabs.harvard.edu/abs/2023ApJ...958..177L}
}

@ARTICLE{Illiano2023,
       author = {{Illiano}, Giulia and {Papitto}, Alessandro and {Sanna}, Andrea and {Bult}, Peter and {Ambrosino}, Filippo and {Miraval Zanon}, Arianna and {Coti Zelati}, Francesco and {Stella}, Luigi and {Altamirano}, Diego and {Baglio}, Maria Cristina and {Bozzo}, Enrico and {Burderi}, Luciano and {de Martino}, Domitilla and {Di Marco}, Alessandro and {di Salvo}, Tiziana and {Ferrigno}, Carlo and {Loktev}, Vladislav and {Marino}, Alessio and {Ng}, Mason and {Pilia}, Maura and {Poutanen}, Juri and {Salmi}, Tuomo},
        title = "{Timing Analysis of the 2022 Outburst of the Accreting Millisecond X-Ray Pulsar SAX J1808.4-3658: Hints of an Orbital Shrinking}",
      journal = {\apjl},
         year = 2023,
        month = jan,
       volume = {942},
       number = {2},
          eid = {L40},
        pages = {L40},
          doi = {10.3847/2041-8213/acad81},
archivePrefix = {arXiv},
       eprint = {2212.09778},
 primaryClass = {astro-ph.HE},
       adsurl = {https://ui.adsabs.harvard.edu/abs/2023ApJ...942L..40I}
}

@ARTICLE{Shakura1973,
       author = {{Shakura}, N.~I. and {Sunyaev}, R.~A.},
        title = "{Black holes in binary systems. Observational appearance.}",
      journal = {\aap},
         year = 1973,
        month = jan,
       volume = {24},
        pages = {337-355},
       adsurl = {https://ui.adsabs.harvard.edu/abs/1973A&A....24..337S}
}

@ARTICLE{Poutanen2009,
       author = {{Poutanen}, Juri and {Ibragimov}, Askar and {Annala}, Marja},
        title = "{On the Nature of Pulse Profile Variations and Timing Noise in Accreting Millisecond Pulsars}",
      journal = {\apjl},
         year = 2009,
        month = nov,
       volume = {706},
       number = {1},
        pages = {L129-L132},
          doi = {10.1088/0004-637X/706/1/L129},
archivePrefix = {arXiv},
       eprint = {0910.5868},
 primaryClass = {astro-ph.HE},
       adsurl = {https://ui.adsabs.harvard.edu/abs/2009ApJ...706L.129P}
}

@ARTICLE{Kajava2011,
       author = {{Kajava}, Jari J.~E. and {Ibragimov}, Askar and {Annala}, Marja and {Patruno}, Alessandro and {Poutanen}, Juri},
        title = "{Varying disc-magnetosphere coupling as the origin of pulse profile variability in SAX J1808.4-3658}",
      journal = {\mnras},
         year = 2011,
        month = oct,
       volume = {417},
       number = {2},
        pages = {1454-1465},
          doi = {10.1111/j.1365-2966.2011.19360.x},
archivePrefix = {arXiv},
       eprint = {1107.0180},
 primaryClass = {astro-ph.HE},
       adsurl = {https://ui.adsabs.harvard.edu/abs/2011MNRAS.417.1454K}
}

@ARTICLE{Petterson1991,
       author = {{Petterson}, J.~A. and {Rothschild}, R.~E. and {Gruber}, D.~E.},
        title = "{A Model for the 35 Day Variations in the Pulse Profile of Hercules X-1}",
      journal = {\apj},
         year = 1991,
        month = sep,
       volume = {378},
        pages = {696},
          doi = {10.1086/170470},
       adsurl = {https://ui.adsabs.harvard.edu/abs/1991ApJ...378..696P}
}

@ARTICLE{Scott2000,
       author = {{Scott}, D. Matthew and {Leahy}, Denis A. and {Wilson}, Robert B.},
        title = "{The 35 Day Evolution of the Hercules X-1 Pulse Profile: Evidence for a Resolved Inner Disk Occultation of the Neutron Star}",
      journal = {\apj},
         year = 2000,
        month = aug,
       volume = {539},
       number = {1},
        pages = {392-412},
          doi = {10.1086/309203},
archivePrefix = {arXiv},
       eprint = {astro-ph/0002327},
 primaryClass = {astro-ph},
       adsurl = {https://ui.adsabs.harvard.edu/abs/2000ApJ...539..392S}
}

@ARTICLE{Zhang2025,
       author = {{Zhang}, Shuang-Nan and {Santangelo}, Andrea and {Xu}, Yupeng and {Feng}, Hua and {Lu}, Fangjun and {Chen}, Yong and {Ge}, Mingyu and {Nandra}, Kirpal and {Wu}, Xin and {Feroci}, Marco and {Hernanz}, Margarita and {Liu}, Congzhan and {He}, Huilin and {Wang}, Yusa and {Jiang}, Weichun and {Cui}, Weiwei and {Yang}, Yanji and {Wang}, Juan and {Li}, Wei and {Li}, Hong and {Du}, Yuanyuan and {Liu}, Xiaohua and {Meng}, Bin and {Wen}, Xiangyang and {Zhang}, Aimei and {Ma}, Jia and {Li}, Maoshun and {Li}, Gang and {Qi}, Liqiang and {Sun}, Jianchao and {Luo}, Tao and {Liu}, Hongwei and {Liu}, Xiaojing and {Zhang}, Fan and {Luo}, Laidan and {Zhu}, Yuxuan and {Zhao}, Zijian and {Sun}, Liang and {Yang}, Xiongtao and {Wu}, Qiong and {Jiang}, Jiechen and {Shi}, Haoli and {Liu}, Jiangtao and {Xu}, Yanbing and {Yang}, Sheng and {Zhang}, Laiyu and {Han}, Dawei and {Gao}, Na and {Huo}, Jia and {Zhang}, Ziliang and {Wang}, Hao and {Zhao}, Xiaofan and {Wang}, Shuo and {Li}, Zhenjie and {Bao}, Ziyu and {Liu}, Yaoguang and {Wang}, Ke and {Wang}, Na and {Wang}, Bo and {Wang}, Langping and {Wang}, Dianlong and {Ding}, Fei and {Sheng}, Lizhi and {Qiang}, Pengfei and {Yan}, Yongqing and {Liu}, Yongan and {Wu}, Zhenyu and {Liu}, Yichen and {Chen}, Hao and {Zhang}, Yacong and {Liu}, Hongbang and {Altmann}, Alexander and {Bechteler}, Thomas and {Burwitz}, Vadim and {Fiorini}, Carlo and {Friedrich}, Peter and {Meidinger}, Norbert and {Strecker}, Rafael and {Baldini}, Luca and {Bellazzini}, Ronaldo and {Bonino}, Raffaella and {Frass{\`a}}, Andrea and {Latronico}, Luca and {Maldera}, Simone and {Manfreda}, Alberto and {Minuti}, Massimo and {Pesce-Rollins}, Melissa and {Sgr{\`o}}, Carmelo and {Tugliani}, Stefano and {Pareschi}, Giovanni and {Basso}, Stefano and {Sironi}, Giorgia and {Spiga}, Daniele and {Tagliaferri}, Gianpiero and {Tykhonov}, Andrii and {Paltani}, St{\`e}phane and {Bozzo}, Enrico and {Tenzer}, Christoph and {Bayer}, J{\"o}rg and {Tuo}, Youli and {Liu}, Honghui and {Zhang}, Yonghe and {Cai}, Zhiming and {Liu}, Huaqiu and {Chen}, Wen and {Wang}, Chunhong and {He}, Tao and {Chen}, Yehai and {Qiu}, Chengbo and {Zhang}, Ye and {Feng}, Jianchao and {Zhu}, Xiaofei and {Zhou}, Heng and {Zheng}, Shijie and {Song}, Liming and {Wang}, Jinzhou and {Jia}, Shumei and {Jiang}, Zewen and {Li}, Xiaobo and {Zhao}, Haisheng and {Guan}, Ju and {Zhang}, Juan and {Li}, Chengkui and {Huang}, Yue and {Liao}, Jinyuan and {You}, Yuan and {Zhang}, Hongmei and {Wang}, Wenshuai and {Wang}, Shuang and {Ou}, Ge and {Hu}, Hao and {Shi}, Jingyan and {Cui}, Tao and {Jiang}, Xiaowei and {Cheng}, Yaodong and {Li}, Haibo and {Xu}, Yanjun and {Zane}, Silvia and {Bambi}, Cosimo and {Bu}, Qingcui and {Dall'Osso}, Simone and {Rosa}, Alessandra De and {Gou}, Lijun and {Guillot}, Sebastien and {Ji}, Long and {Li}, Ang and {Mao}, Jirong and {Patruno}, Alessandro and {Stratta}, Giulia and {Taverna}, Roberto and {Tsygankov}, Sergey and {Uttley}, Phil and {Watts}, Anna L. and {Wu}, Xuefeng and {Xu}, Renxin and {Yi}, Shuxu and {Zhang}, Guobao and {Zhang}, Liang and {Zhao}, Wen and {Zhou}, Ping},
        title = "{The enhanced X-ray Timing and Polarimetry mission{\textemdash}eXTP for launch in 2030}",
      journal = {Science China Physics, Mechanics, and Astronomy},
         year = 2025,
        month = sep,
       volume = {68},
       number = {11},
          eid = {119502},
        pages = {119502},
          doi = {10.1007/s11433-025-2786-6},
archivePrefix = {arXiv},
       eprint = {2506.08101},
 primaryClass = {astro-ph.HE},
       adsurl = {https://ui.adsabs.harvard.edu/abs/2025SCPMA..6819502Z}
}

@ARTICLE{Illarionov1975,
       author = {{Illarionov}, A.~F. and {Sunyaev}, R.~A.},
        title = "{Why the Number of Galactic X-ray Stars Is so Small?}",
      journal = {\aap},
         year = 1975,
        month = feb,
       volume = {39},
        pages = {185},
       adsurl = {https://ui.adsabs.harvard.edu/abs/1975A&A....39..185I}
}

@ARTICLE{Kini24,
       author = {{Kini}, Yves and {Salmi}, Tuomo and {Vinciguerra}, Serena and {Watts}, Anna L. and {Bilous}, Anna and {Galloway}, Duncan K. and {van der Wateren}, Emma and {Khalsa}, Guru Partap and {Bogdanov}, Slavko and {Buchner}, Johannes and {Suleimanov}, Valery},
        title = "{Constraining the properties of the thermonuclear burst oscillation source XTE J1814-338 through pulse profile modelling}",
      journal = {\mnras},
         year = 2024,
        month = dec,
       volume = {535},
       number = {2},
        pages = {1507-1525},
          doi = {10.1093/mnras/stae2398},
archivePrefix = {arXiv},
       eprint = {2405.10717},
 primaryClass = {astro-ph.HE},
       adsurl = {https://ui.adsabs.harvard.edu/abs/2024MNRAS.535.1507K}
}

@ARTICLE{Kini26,
       author = {{Kini}, Yves and {Mauviard}, Lucien and {Salmi}, Tuomo and {Watts}, Anna L. and {Guillot}, Sebastien and {Dorsman}, Bas and {Choudhury}, Devarshi and {Gonz{\'a}lez-Caniulef}, Denis and {Hoogkamer}, Mariska and {Huppenkothen}, Daniela and {Kazantsev}, Christine and {Kerr}, Matthew and {Nissanke}, Samaya and {Ray}, Paul S. and {Stammler}, Pierre and {Vinciguerra}, Serena},
        title = "{A NICER View of PSR J0030+0451: Updated Constraints from Six Years of NICER Observations}",
      journal = {arXiv e-prints},
         year = 2026,
        month = feb,
          eid = {arXiv:2602.23743},
        pages = {arXiv:2602.23743},
          doi = {10.48550/arXiv.2602.23743},
archivePrefix = {arXiv},
       eprint = {2602.23743},
 primaryClass = {astro-ph.HE},
       adsurl = {https://ui.adsabs.harvard.edu/abs/2026arXiv260223743K}
}

@ARTICLE{Rappaport2004,
       author = {{Rappaport}, S.~A. and {Fregeau}, J.~M. and {Spruit}, H.},
        title = "{Accretion onto Fast X-Ray Pulsars}",
      journal = {\apj},
         year = 2004,
        month = may,
       volume = {606},
       number = {1},
        pages = {436-443},
          doi = {10.1086/382863},
archivePrefix = {arXiv},
       eprint = {astro-ph/0310224},
 primaryClass = {astro-ph},
       adsurl = {https://ui.adsabs.harvard.edu/abs/2004ApJ...606..436R}
}

@ARTICLE{Papitto2015,
       author = {{Papitto}, A. and {Torres}, D.~F.},
        title = "{A Propeller Model for the Sub-luminous State of the Transitional Millisecond Pulsar PSR J1023+0038}",
      journal = {\apj},
         year = 2015,
        month = jul,
       volume = {807},
       number = {1},
          eid = {33},
        pages = {33},
          doi = {10.1088/0004-637X/807/1/33},
archivePrefix = {arXiv},
       eprint = {1504.05029},
 primaryClass = {astro-ph.HE},
       adsurl = {https://ui.adsabs.harvard.edu/abs/2015ApJ...807...33P}
}

@ARTICLE{Romanova2004,
       author = {{Romanova}, M.~M. and {Ustyugova}, G.~V. and {Koldoba}, A.~V. and {Lovelace}, R.~V.~E.},
        title = "{Three-dimensional Simulations of Disk Accretion to an Inclined Dipole. II. Hot Spots and Variability}",
      journal = {\apj},
         year = 2004,
        month = aug,
       volume = {610},
       number = {2},
        pages = {920-932},
          doi = {10.1086/421867},
archivePrefix = {arXiv},
       eprint = {astro-ph/0404496},
 primaryClass = {astro-ph},
       adsurl = {https://ui.adsabs.harvard.edu/abs/2004ApJ...610..920R}
}

@ARTICLE{Romanova2005,
       author = {{Romanova}, M.~M. and {Ustyugova}, G.~V. and {Koldoba}, A.~V. and {Lovelace}, R.~V.~E.},
        title = "{Propeller-driven Outflows and Disk Oscillations}",
      journal = {\apjl},
         year = 2005,
        month = dec,
       volume = {635},
       number = {2},
        pages = {L165-L168},
          doi = {10.1086/499560},
archivePrefix = {arXiv},
       eprint = {astro-ph/0511600},
 primaryClass = {astro-ph},
       adsurl = {https://ui.adsabs.harvard.edu/abs/2005ApJ...635L.165R}
}

@ARTICLE{Gandolfi2012,
       author = {{Gandolfi}, S. and {Carlson}, J. and {Reddy}, Sanjay},
        title = "{Maximum mass and radius of neutron stars, and the nuclear symmetry energy}",
      journal = {\prc},
         year = 2012,
        month = mar,
       volume = {85},
       number = {3},
          eid = {032801},
        pages = {032801},
          doi = {10.1103/PhysRevC.85.032801},
archivePrefix = {arXiv},
       eprint = {1101.1921},
 primaryClass = {nucl-th},
       adsurl = {https://ui.adsabs.harvard.edu/abs/2012PhRvC..85c2801G}
}

@article{Kass1995,
author = {Robert E. Kass and Adrian E. Raftery},
title = {Bayes Factors},
journal = {Journal of the American Statistical Association},
volume = {90},
number = {430},
pages = {773--795},
year = {1995},
publisher = {Taylor \& Francis},
doi = {10.1080/01621459.1995.10476572},
URL = {https://doi.org/10.1080/01621459.1995.10476572},
eprint = {https://doi.org/10.1080/01621459.1995.10476572}
}

@ARTICLE{Holt2025,
       author = {{Holt}, Isiah M. and {Miller}, M. Coleman and {Dittmann}, Alexander J. and {Lamb}, Frederick K.},
        title = "{An Investigation of Systematic Effects from Background Priors on PSR J0740$+$6620 Radius Estimates using Synthetic NICER and XMM-Newton Data}",
      journal = {arXiv e-prints},
         year = 2025,
        month = nov,
          eid = {arXiv:2511.16759},
        pages = {arXiv:2511.16759},
          doi = {10.48550/arXiv.2511.16759},
archivePrefix = {arXiv},
       eprint = {2511.16759},
 primaryClass = {astro-ph.HE},
       adsurl = {https://ui.adsabs.harvard.edu/abs/2025arXiv251116759H}
}

@PHDTHESIS{Riley2019PhD,
       author = {{Riley}, Thomas Edward},
        title = "{Neutron star parameter estimation from a NICER perspective}",
       school = {University of Amsterdam, Netherlands},
         year = 2019,
        month = jan,
       adsurl = {https://ui.adsabs.harvard.edu/abs/2019PhDT........97R}
}

@ARTICLE{Choudhury24b,
       author = {{Choudhury}, Devarshi and {Watts}, Anna L. and {Dittmann}, Alexander J. and {Miller}, M. Coleman and {Morsink}, Sharon M. and {Salmi}, Tuomo and {Vinciguerra}, Serena and {Bogdanov}, Slavko and {Guillot}, Sebastien and {Wolff}, Michael T. and {Arzoumanian}, Zaven},
        title = "{Exploring Waveform Variations among Neutron Star Ray-tracing Codes for Complex Emission Geometries}",
      journal = {\apj},
         year = 2024,
        month = nov,
       volume = {975},
       number = {2},
          eid = {202},
        pages = {202},
          doi = {10.3847/1538-4357/ad7255},
archivePrefix = {arXiv},
       eprint = {2406.07285},
 primaryClass = {astro-ph.HE},
       adsurl = {https://ui.adsabs.harvard.edu/abs/2024ApJ...975..202C}
}

@ARTICLE{Galloway2008,
       author = {{Galloway}, Duncan K. and {Muno}, Michael P. and {Hartman}, Jacob M. and {Psaltis}, Dimitrios and {Chakrabarty}, Deepto},
        title = "{Thermonuclear (Type I) X-Ray Bursts Observed by the Rossi X-Ray Timing Explorer}",
      journal = {\apjs},
         year = 2008,
        month = dec,
       volume = {179},
       number = {2},
        pages = {360-422},
          doi = {10.1086/592044},
archivePrefix = {arXiv},
       eprint = {astro-ph/0608259},
 primaryClass = {astro-ph},
       adsurl = {https://ui.adsabs.harvard.edu/abs/2008ApJS..179..360G}
}

@ARTICLE{Dorsman2026,
       author = {{Dorsman}, Bas and {Salmi}, Tuomo and {Watts}, Anna L. and {Ng}, Mason and {Bobrikova}, Anna and {Di Marco}, Alessandro and {Galloway}, Duncan K. and {Guillot}, Sebastien and {Hoogkamer}, Mariska and {Kini}, Yves and {La Monaca}, Fabio and {Loktev}, Vladislav and {Lucchini}, Matteo and {Malacaria}, Christian and {Mao}, Ying-Han and {Papitto}, Alessandro and {Poutanen}, Juri},
        title = "{Pulse profile modelling of the 2024 outburst of the accreting millisecond pulsar SRGA J144459.2-604207}",
      journal = {arXiv e-prints},
         year = 2026,
        month = may,
          eid = {arXiv:2605.18731},
        pages = {arXiv:2605.18731},
          doi = {10.48550/arXiv.2605.18731},
archivePrefix = {arXiv},
       eprint = {2605.18731},
 primaryClass = {astro-ph.HE},
       adsurl = {https://ui.adsabs.harvard.edu/abs/2026arXiv260518731D}
}



\appendix


\bsp	
\label{lastpage}
\end{document}